%% file: main.tex
\documentclass[sigplan,screen,pbalance=true]{acmart}

\copyrightyear{2026}
\acmYear{2026}
\setcopyright{cc}
\setcctype{by}
\acmConference[ASPLOS '26]{Proceedings of the 31st ACM International Conference on Architectural Support for Programming Languages and Operating Systems, Volume 2}{March 22--26, 2026}{Pittsburgh, PA, USA}
\acmBooktitle{Proceedings of the 31st ACM International Conference on Architectural Support for Programming Languages and Operating Systems, Volume 2 (ASPLOS '26), March 22--26, 2026, Pittsburgh, PA, USA}
\acmPrice{}
\acmDOI{10.1145/3779212.3790236}
\acmISBN{979-8-4007-2359-9/2026/03}

\input{section/environment}
\newcommand{\sysname}{{M\textsc{ux}W\textsc{ise}}}

\begin{document}

%
%
\title[Prefill-decode Multiplexing]{Towards High-Goodput LLM Serving with Prefill-decode Multiplexing}


\author{Yukang Chen}
\authornote{Equal contribution.}
\email{chenyukang@sjtu.edu.cn}
\orcid{0009-0007-7719-1377}
\affiliation{%
  \institution{Shanghai Jiao Tong University}
  \city{Shanghai}
  \country{China}
}

\author{Weihao Cui}
\authornotemark[1]
\orcid{0000-0002-6646-5260}
\email{weihao@sjtu.edu.cn}
\affiliation{%
  \institution{Shanghai Jiao Tong University}
  \city{Shanghai}
  \country{China}
}
\affiliation{%
  \institution{National University of Singapore}
  \country{Singapore}
}

\author{Han Zhao}
\authornotemark[1]
\orcid{0000-0002-1561-5329}
\email{zhao-han@cs.sjtu.edu.cn}
\affiliation{%
  \institution{Shanghai Jiao Tong University}
  \city{Shanghai}
  \country{China}
}

\author{Ziyi Xu}
\orcid{0009-0000-4411-9773}
\email{xzy2022@sjtu.edu.cn}
\affiliation{%
  \institution{Shanghai Jiao Tong University}
  \city{Shanghai}
  \country{China}
}

\author{Xiaoze Fan}
\orcid{0009-0004-8742-606X}
\email{jasonfxz@sjtu.edu.cn}
\affiliation{%
  \institution{Shanghai Jiao Tong University}
  \city{Shanghai}
  \country{China}
}

\author{Xusheng Chen}
\orcid{0000-0002-2807-9780}
\email{michael.xschen@gmail.com}
\affiliation{%
  \institution{Researcher}
  \city{Shanghai}
  \country{China}
}

\author{Yangjie Zhou}
\orcid{0000-0002-3652-5437}
\email{yj_zhou@nus.edu.sg}
\affiliation{%
  \institution{National University of Singapore}
  \country{Singapore}
}

\author{Shixuan Sun}
\orcid{0000-0003-4060-9438}
\email{sunshixuan@sjtu.edu.cn}
\affiliation{%
  \institution{Shanghai Jiao Tong University}
  \city{Shanghai}
  \country{China}
}

\author{Bingsheng He}
\orcid{0000-0001-8618-4581}
\email{dcsheb@nus.edu.sg}
\affiliation{%
  \institution{National University of Singapore}
  \country{Singapore}
}

\author{Quan Chen}
\authornote{Corresponding author.}
\orcid{0000-0001-5832-0347}
\email{chen-quan@cs.sjtu.edu.cn}
\affiliation{%
  \institution{Shanghai Jiao Tong University}
  \city{Shanghai}
  \country{China}
}
\renewcommand{\shortauthors}{Yukang Chen et al.}

\begin{abstract}

Large Language Model (LLM) serving must meet stringent Service Level Objectives (SLOs) for both the prefill and decode phases.
Some existing solutions disaggregate the two phases, causing potential resource idleness or compute redundancy.
Others split the prefill phase into chunks and fuse it with decode iteration, creating a dilemma between SLO compliance and high utilization.
To address these issues, an efficient serving system should dynamically adapt compute allocation, decouple compute from memory management, and execute prefill and decode independently.
We present \sysname{}, an LLM serving framework that adopts a new paradigm, intra-GPU prefill-decode multiplexing, to meet these requirements.
To fully exploit the paradigm, \sysname{} integrates a bubble-less multiplex engine, a contention-tolerant estimator, and an SLO-aware dispatcher.
Evaluation shows that \sysname{} improves peak throughput under SLO guarantees by an average of $2.20\times$ (up to $3.06\times$) over state-of-the-art baselines.
\end{abstract}




\maketitle

\input{section/introduction}

\input{section/background}

\input{section/overview}

\input{section/pooling_compute}

\input{section/evaluation}
\input{section/discussion}

\input{section/related_work}

\input{section/conclusion}

\begin{acks}
This work is partially sponsored by the National Key Research and Development Program of China (2024YFB4505700), National Natural Science Foundation of China (62232011) and Natural Science Foundation of Shanghai Municipality (24ZR1430500).
We thank the anonymous reviewers for their constructive feedback and suggestions.
\end{acks}

\input{section/ae}

\bibliographystyle{ACM-Reference-Format}
\bibliography{reference.bib}

\end{document}

%% file: section/environment.tex
\usepackage{balance}
\usepackage{booktabs}
\usepackage{tikz}
\usepackage{comment}
\usepackage{amsmath}
\usepackage{caption}
\usepackage{subcaption}
\usepackage{graphicx}
\usepackage{float}
\usepackage{algorithm}
\usepackage{algorithmicx}
\usepackage[noend]{algpseudocode}
\algnewcommand{\LineComment}[1]{\State \(//\) #1}
\usepackage[normalem]{ulem}
\usepackage{bbding}
\usepackage{todonotes}
\usepackage{filecontents}
\usepackage[cachedir=.]{minted}
\usepackage{enumitem}
\usepackage[bottom]{footmisc}
\usepackage[]{hyperref}
\usepackage{cleveref}
\usepackage{pifont}
\usepackage{multirow}
\usepackage{xcolor}
\usepackage{xparse}
\usepackage{outlines}
\usepackage{lipsum}

\definecolor{LightGray}{gray}{0.9}
\definecolor{FadedBanana}{RGB}{255,255,191}
\definecolor{DeepChalk}{RGB}{255,191,191}
\definecolor{FadedFlora}{RGB}{191,255,191}
\definecolor{DeepSnow}{RGB}{191,255,255}
\definecolor{SoapStone}{RGB}{218,218,218}
\definecolor{LightCayenneSixty}{RGB}{239,206,211}
\definecolor{LightCayenne}{RGB}{208,143,145}

\algrenewcommand\algorithmicrequire{\textbf{Input:}}
\algrenewcommand\algorithmicensure{\textbf{Output:}}

\newcommand{\mypar}[1]{
\vspace{0.5ex}%
\noindent\hspace*{1em}\textbf{\textit{#1. }}}%

%% file: section/introduction.tex
\section{Introduction}

Large language models (LLM) services now perform well across diverse workloads~\cite{openai2025chatgpt,qiMutualReasoning,cursor2025}.
At the request level, an LLM processes input in two phases: a prefill phase that produces the first token, followed by a decode phase that iteratively generates the remaining tokens.
The ratio of input length (prefill) to output length (decode) varies across tasks~\cite{bai2024longbench2,weiChainofthoughtPrompting}.
At the application level, tasks such as chatbot services or agent-based workloads~\cite{qinMooncakeTrading} often consist of multiple turns of requests with shared context.

\begin{figure}
    \centering
    \includegraphics[width=.9\linewidth]{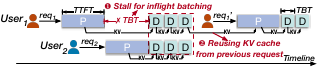}
    \caption{A typical workflow in LLM serving systems: \texttt{User1} sends two consecutive requests, with the second reusing the context from the first. \texttt{User2} sends 1 request.}
    \label{fig:llm-workflow}
\end{figure}

To achieve high throughput for serving these workloads, existing LLM serving systems employ several optimizations.
\autoref{fig:llm-workflow} presents a typical workflow.
While requests arrive at different times, inflight batching stalls the ongoing decode phase to prefill new requests and then processes all decode iterations together in a single batch.
It greatly improves compute utilization for the memory-intensive decode phase~\cite{yuOrcaDistributed}.
Since multi-turn requests share context, LLM serving systems reuse intermediate results (i.e., the KV cache) both within and across requests through a KV cache pool~\cite{zhengEfficientlyProgramming,linParrotEfficient}.

LLM services also impose stringent Service Level Objectives (SLOs).
For instance, chatbot typically requires Time-To-First-Token (TTFT) under 500 ms for prefill and Time-Between-Tokens (TBT) under 100 ms for decode.
Since prefill and decode interleave in an LLM serving system, SLO violations may arise.
In \autoref{fig:llm-workflow}, inflight batching stalls ongoing decode.
A long prefill can thus delay decode, potentially violating its SLO.
To sustain high goodput--peak throughput with SLO guarantees--existing methods fall into two categories: disaggregated serving~\cite{zhongDistServeDisaggregating,patelSplitwiseEfficient} and chunked prefill~\cite{agrawalTamingThroughputlatency}.

As for disaggregated serving, Splitwise~\cite{patelSplitwiseEfficient} separates the prefill and decode phases into distinct instances for SLO guarantees, which has two drawbacks.
Firstly, it cannot adapt to serving dynamics.
In Splitwise, GPUs are statically allocated at initialization.
Under fluctuating request loads and diverse serving patterns, it often leads to resource underutilization.
Secondly, it decreases goodput due to shrinking the KV cache pool.
With the same number of GPUs, disaggregation allocates a separate cache pool for each instance, reducing the effective cache pool size.
This lowers cache hit rate~\cite{yaoCacheBlendFast} (e.g., from $36.6\%$ to $4.2\%$), leading to unnecessary recomputation and degraded goodput.
Furthermore, while LoongServe~\cite{wuLoongServeEfficiently} supports dynamic GPU allocation based on the request sequence length and execution phase, it cannot support the cross-request KV cache reuse, incurring significant recomputation overhead in multi-turn workloads.

Chunked-prefill~\cite{agrawalTamingThroughputlatency} is another approach to meet decode SLOs.
It splits the prefill phase into chunks within each GPU and fuses each chunk with a decode iteration.
To ensure computational equivalence, each chunk reads the KV cache generated by all previous chunks.
It ensures the decode SLOs by capping the token budget, defined as the sum of new tokens from the prefill chunk and the decode batch.
By tuning the chunk and decode batch sizes, it adapts to serving dynamics.
Since it avoids disaggregation, it also prevents goodput loss from a reduced KV cache pool.

Unfortunately, chunking is not a free lunch.
It creates a dilemma between SLO compliance and high utilization.
Because prefill chunk and decode iteration must execute together, the token budget governs both decode SLO attainment and GPU saturation.
Yet, finding a sweet budget in practice is infeasible.
E.g., deploying a 70B LLM on 8 A100 GPUs requires a 4K budget to saturate the GPU, which is $8\times$ larger than the SLO-compliant budget (256 for a 100 ms TBT SLO).
Moreover, TBT in chunk-prefill is inflated by repetitive KV cache access from the prefill chunk.
With extremely long reused context, common in multi-turn workloads, chunked-prefill may even fail to meet SLO guarantees (\S\ref{sec:moti-chunked}).
Ultimately, chunked-prefill cannot sustain high goodput.

Achieving high-goodput LLM serving requires more flexible compute management.
We propose intra-GPU prefill-decode (PD) multiplexing as a promising new serving paradigm.
Specifically, the prefill and decode phases are executed on different streaming multiprocessors (SMs) within the GPUs.
In the new paradigm,
1) compute partitions can be reconfigured with low overhead to adapt to serving dynamics;
2) multiplexed phases share GPU memory, keeping the KV cache pool efficient;
3) with spatial sharing, prefill and decode execute independently, avoiding the tradeoff between SLO compliance and utilization.

Realizing this paradigm is non-trivial.
Firstly, phase coordination is still required to enable inflight batching and improve compute utilization.
However, since prefill and decode latencies differ significantly, naive coordination often leaves GPU bubbles.
Secondly, spatial multiplexing introduces unpredictable contention.
Although existing approaches~\cite{nvidia2025mig,nvidia_mps,nvidia2025grecontexts} partition compute, they provide little control over shared resources such as memory bandwidth.

To this end, we propose \sysname{}, an LLM serving framework that achieves high goodput across diverse workloads.
\sysname{} comprises three modules: a \textit{bubble-less multiplex engine}, a \textit{contention-tolerant estimator}, and an \textit{SLO-aware dispatcher}.
The engine partitions prefill into layers with negligible overhead, aligning execution latencies for bubble-less multiplexing.
The contention-tolerant estimator provides worst-case latency predictions by combining a solo-run predictor with a contention guard derived from one-time offline profiling.
Built atop the engine and estimator, the SLO-aware dispatcher schedules diverse LLM requests efficiently by selecting multiplexing plans to maximize goodput.

We implement \sysname{} on top of SGLang~\cite{zhengEfficientlyProgramming}, extending it with PD multiplexing.
\sysname{} is evaluated extensively on both small and large LLMs using real-world workloads.
Experiments show that \sysname{} achieves an average $2.20\times$ goodput improvement (up to $3.06\times$) over state-of-the-art solutions.
In summary, our contributions are:
\begin{itemize}
[leftmargin=*,topsep=0.2em]
\item We identify key requirements for LLM serving with high goodput through a detailed analysis of prior works.
\item We propose a new LLM serving paradigm--PD multiplexing--aligned with these requirements, and present a clean design to effectively serve LLMs with high goodput.
\item We evaluate \sysname{} under diverse workloads, demonstrating its superiority over state-of-the-art solutions.
\end{itemize}

%% file: section/background.tex
\section{Background \& Motivation}

\subsection{LLM Services}\label{sec:llm-services}

\begin{figure}
    \centering
    \includegraphics[width=0.65\linewidth]{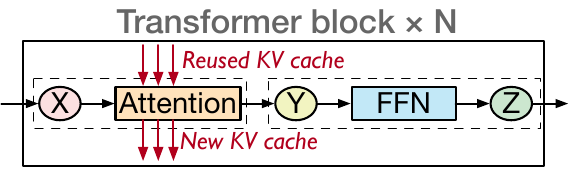}
    \caption{Main architecture of most LLMs.}
    \label{fig:llm-arch}
\end{figure}

\mypar{Architecture of LLMs}\label{sec:llm-arch}
Most LLMs~\cite{dubeyLlama3,touvronLlama2,touvronLLaMAOpen,baiQwenTechnical} are built upon the transformer architecture~\cite{vaswaniAttentionAll}, with model-specific modifications.
\autoref{fig:llm-arch} illustrates a typical transformer layer, which is replicated multiple times to form an LLM model.
Each transformer layer contains an attention layer and a feed-forward network (FFN) layer.

Attention computation requires access to all keys and values from processed tokens and also generates the keys and values of new tokens.
To avoid redundant computation, LLM serving systems store this data in a KV cache. In the prefill phase, the KV cache is populated from the requests in previous turns. In each decode iteration, the KV cache is derived from earlier prefill and decode iterations.

\begin{table}
\caption{Diverse patterns of typical LLM tasks. Minimum, mean, and maximum values for each metric are reported. The input length includes the length of new and reused context.}
\label{tb:tasks}
\setlength{\tabcolsep}{2pt}
\centering
\footnotesize
\begin{tabular}{c|ccc}
\toprule
 & \textbf{Input length} & \textbf{Output length} & \textbf{Reused length}    \\
\midrule
\textbf{ShareGPT~\cite{sharegpt_2023}}     & 4/226/1024       & 4/195/1838    & \textbackslash{} \\
\textbf{LooGLE~\cite{li2024looglelongcontextlanguagemodels}}      & 3380/30k/81k & 2/15/326      & \textbackslash{} \\
\textbf{OpenThoughts~\cite{guha_openthoughts_2025}} & 311/709/4633            & 684/8374/32k         & 243 \\
\textbf{Conversation~\cite{qinMooncakeTrading}} & 891/7538/123k    & 1/342/2000    & 0/4496/120k    \\
\textbf{Tool\&agent~\cite{qinMooncakeTrading}}  & 891/8596/123k    & 1/182/2000    & 0/4905/120k   \\
\bottomrule
\end{tabular}
\end{table}

\begin{figure}
    \centering
    \includegraphics[width=.95\linewidth]{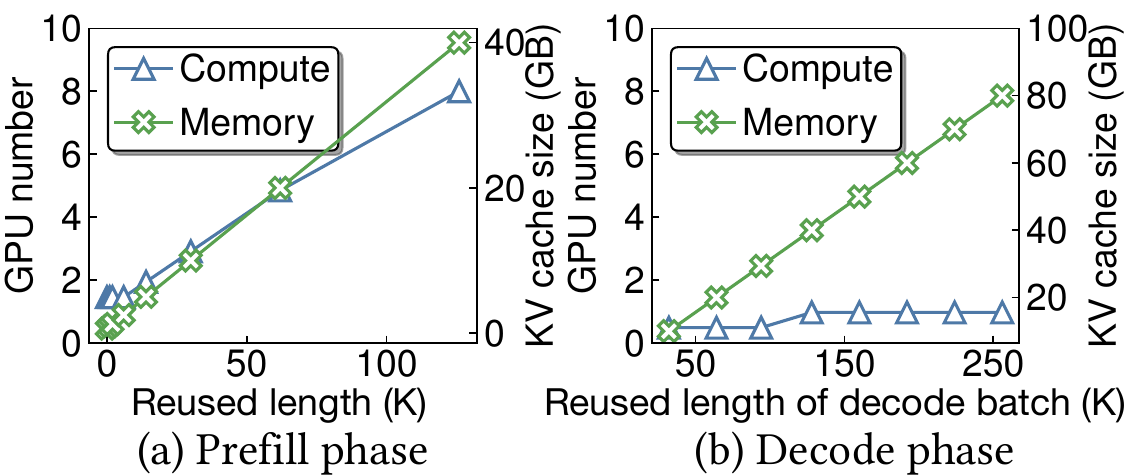}
    \caption{Required compute and memory for processing different phases under SLO constraints with varied reused context lengths.
    For prefill (a), the batch size is fixed at $1$, the new context length is set to $2K$, and TTFT is set to $400ms$. For decode (b), the batch size is fixed at $32$, and TBT is set to $100ms$. These settings are commonly seen in online serving.}
    \label{fig:pooling-requirement}
\end{figure}

\begin{figure*}
\centering
    \begin{minipage}[t]{0.72\textwidth}
    \centering
    \includegraphics[width=\linewidth]{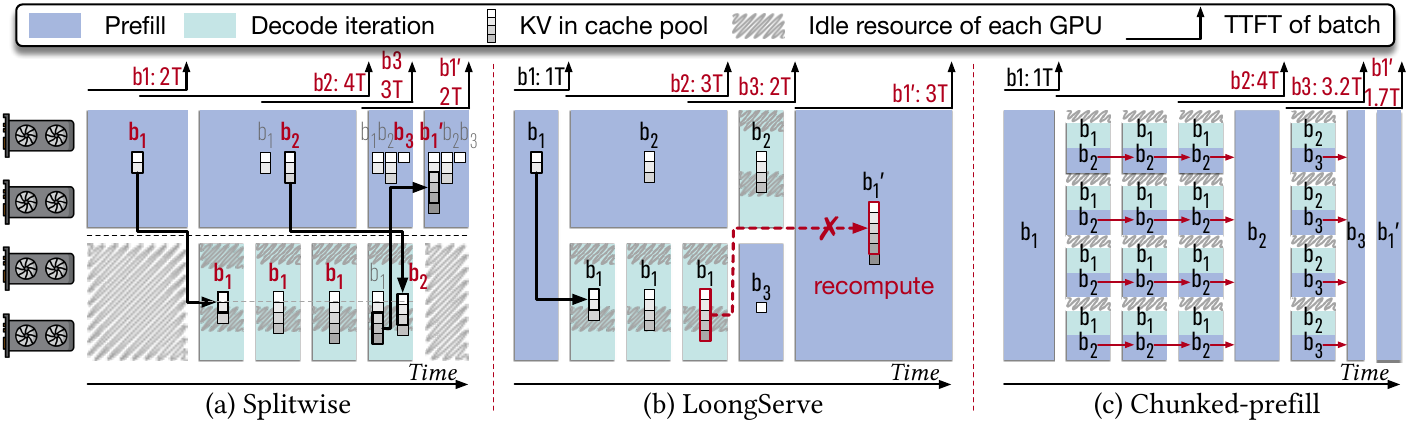}
    \captionof{figure}{
    Processing four LLM request batches on 4 GPUs using (a) Splitwise, (b) LoongServe (c) chunked-prefill.
    All methods satisfy the TBT SLO (T per decode iteration).
    Specifically, $b_1$ arrives at $0T$, $b_2$ at $1T$, $b_3$ at $3T$, and $b_1’$ at $5T$.
    $b_1’$ denotes a subsequent request batch that reuses the KV cache of $b_1$.
    Inefficient TTFTs are marked in red for each method.
    KV cache management is shown only for the two disaggregated methods, as they require migration or recomputation.
    In (a) and (b), solid black arrows represent migration, while dashed red arrows with cross markers denote recomputation.
    In (a), the KV cache column with a red $b_i$ indicates the active batch.
    In (c), the red arrow denotes KV cache reads from earlier chunks.
    }
    \label{fig:llm-serving}
\end{minipage}
\hspace{1mm}
    \begin{minipage}[t]{0.25\textwidth}
    \centering
   \includegraphics[width=\linewidth]{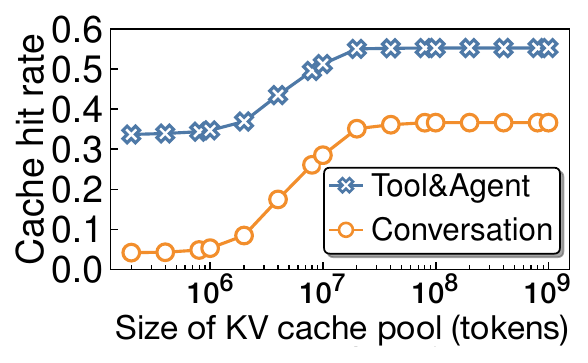}
    \captionof{figure}{Cache hit rates under varying capacities of the KV cache pool. The eviction policy is Least Recently Used. For serving a 70B LLM, achieving the optimal hit rate requires 3.3 TB of memory. Workload trace details are shown in \autoref{tb:tasks}.}
    \label{fig:cache_hit}
\end{minipage}
\end{figure*}

\mypar{Diverse workload patterns}
\autoref{tb:tasks} illustrates the diverse patterns of five typical LLM tasks.
The first three are single-turn requests: ShareGPT~\cite{sharegpt_2023} is a chatbot task, LooGLE~\cite{li2024looglelongcontextlanguagemodels} is a long-context understanding task, and OpenThoughts~\cite{guha_openthoughts_2025} is a reasoning task.
LooGLE has a long input length due to long documents.
Reasoning often requires long thought processes, so OpenThoughts tends to have a longer output length than others.
Requests in OpenThoughts share the same system prompt, which is a constant input context (i.e., reused length in the table).
Conversion and Tool\&agent~\cite{qinMooncakeTrading} are two real-world multi-turn tasks.
The output tokens from earlier requests become the input context for later requests in the same session.
We use these workloads to conduct experiments that both motivate and evaluate our design.

\subsection{Characterization under SLO constraint}
\label{sec:exp_analysis}
Many prior works~\cite{zhongDistServeDisaggregating, patelSplitwiseEfficient, wuLoongServeEfficiently} have investigated the relationship between resource requirements and SLO attainment concerning input length and batch size.
Their experiments show that the prefill phase is compute-intensive, with compute demand growing linearly with input length, while the decode phase is memory-intensive.
However, they mainly focus on the simple single-turn case, which does not consider the effect of reused input length.

Under these circumstances, we further study how the reused length impacts the compute and memory demands of prefill and decode. In our experiment, the reused length spans the range shown in \autoref{tb:tasks}, and LlaMA-70B~\cite{dubeyLlama3} is deployed with tensor parallelism~\cite{zhengAlpaAutomating} on a server with 8 A100 GPUs. All GPUs are configured with the same partial compute resource, defined by the SM number. For each reused length, we determine the best-fit GPU partition ratio (denoted as $GPU_{ratio}$) to satisfy the SLO target. \autoref{fig:pooling-requirement} reports the total compute demand of LlaMA-70B under different reused lengths, computed as $GPU_{num} = GPU_{ratio} \times 8$.

As shown in \autoref{fig:pooling-requirement}-(a), prefill phase requires increasingly more compute resources to meet SLO targets as the reused length grows.
In contrast, the compute demand of the decode phase shows less sensitivity.
Thus, it is also critical to allocate more compute to the prefill phase as the reused length increases.
Further, the distinct compute requirements of two phases necessitate a runtime compute resource partition for SLO attainment and high utilization.

\autoref{fig:pooling-requirement}-(b) shows that the KV cache required by both the prefill and decode easily reaches tens or even hundreds of gigabytes.
This is common in multi-turn LLM services, which produce ultra-long reused contexts.
It is preferable to keep the KV cache in the same memory space (aggregated serving) for efficient reuse across phases and requests.

In a nutshell, we make two observations:
\textit{1) Appropriate and dynamic compute partition is essential for meeting the distinct SLO targets of different phases under diverse workloads.
2) Reusing the KV cache across phases and requests is critical for reducing redundant computation and improving goodput.}

\subsection{Deficiencies of Existing Works}
\label{sec:deficiencies}

\subsubsection{Disaggregated Serving}
Disaggregating approaches partition GPUs across phases to meet the SLO targets in LLM serving and can be further divided into static and dynamic disaggregation methods. \autoref{fig:llm-serving}-(a) illustrates the static approach (Splitwise~\cite{patelSplitwiseEfficient}), while \autoref{fig:llm-serving}-(b) shows the dynamic approach (LoongServe~\cite{wuLoongServeEfficiently}).

\mypar{Static disaggregation}
As shown in \autoref{fig:llm-serving}-(a), there is a prefill instance and a decode instance with Splitwise~\cite{patelSplitwiseEfficient}. Each instance occupies two GPUs statically and has its own KV cache pool. The GPU number is static after the instance is initialized. In this case, Splitwise suffers from two problems.

\textit{First, Splitwise does not adapt to serving dynamics.}
For example, when batch \texttt{b1} arrives, only two GPUs process the prefill while the other two GPUs for decoding remain idle. In online serving, such idle periods are common as request loads fluctuate. \textit{Second, the coupled management of compute and memory introduces further inefficiencies.}
For instance, if the decode phase of \texttt{b1} in \autoref{fig:llm-serving}-(a) requires two GPUs to store the KV cache, the system must also allocate two GPUs for computation.
Since compute and memory requirements are misaligned, as shown in \autoref{fig:pooling-requirement}-(b), the GPUs’ compute resources may be underutilized.

In addition, each instance must maintain its own model weights and KV cache pool. As a result, the KV cache pool in \autoref{fig:llm-serving}-(a) is at most half the size of that with four GPUs under non-disaggregated execution. Furthermore, experimental results in \autoref{fig:cache_hit} show that this reduced capacity sharply lowers the KV cache hit rate in multi-turn workloads, ultimately degrading the system's goodput.

\mypar{Dynamic disaggregation}
LoongServe~\cite{wuLoongServeEfficiently} supports dynamic GPU partitioning across the two phases. Specifically, it scales GPU resources based on the sequence length and execution phase. As shown in \autoref{fig:llm-serving}-(b), when batch b1 arrives, the scheduler assigns four GPUs to prefill. After prefill, it scales down to two GPUs for the decode iterations.

However, LoongServe still causes idleness due to coupled management, and worse, it trades KV cache reuse for adaptiveness needed in serving dynamics. To avoid duplication, it immediately releases the KV cache on original GPUs. Thus, KV caches are reused only from prefill to decode within a single request and cannot be reused across multi-turn requests. In \autoref{fig:llm-serving}-(b), when \texttt{b1’} needs to reuse the KV cache generated by \texttt{b1}, LoongServe recomputes the entire KV cache.

\begin{figure}
    \centering
    \includegraphics[width=.95\linewidth]{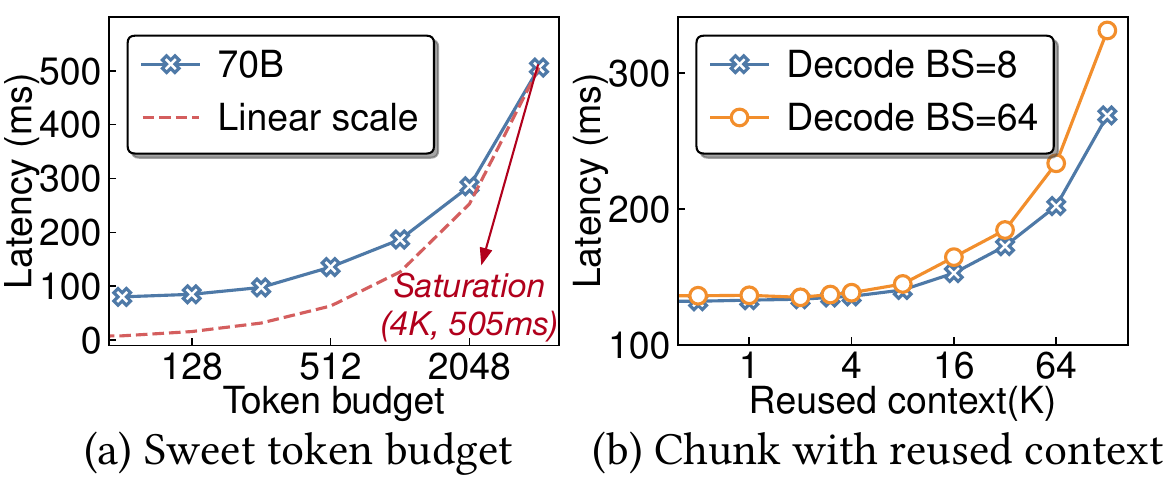}
    \caption{(a) Sweet spot of the token budget in chunk-prefill. The decode uses a fixed batch size of 32, with each request having a reused context length of 1K tokens. (b) Latencies with varied reused context of the fused prefill chunk in chunk-prefill. The token budget is fixed at 512, and the reused context length of decode phase is the same as in (a).
    }
    \label{fig:sweet_spot}
\end{figure}

\subsubsection{Chunked-prefill}
\label{sec:moti-chunked}
Chunked-prefill~\cite{agrawalTamingThroughputlatency} adopts intra-GPU compute fusion.
As shown in \autoref{fig:llm-serving}-(c), it splits prefill into chunks and fuses each chunk with a decode iteration.
To guarantee decode SLOs, chunked-prefill caps the token budget, which is the sum of new tokens from the prefill chunk and the decode batch.
While chunked-prefill has known drawbacks such as quadratic memory overhead~\cite{zhongDistServeDisaggregating}, we find another drawback. Specifically, chunking introduces a dilemma between SLO attainment and utilization.

\autoref{fig:sweet_spot}-(a) presents TBT in Chunked-prefill of varying token budget. In this experiment, the decode iteration for fusion has a static batch size of 32 and a reused context length of 1K tokens, and Llama3-70B is deployed on a server with 8 A100 GPUs. As shown, the latency does not increase linearly with the token budget until it reaches $4K$. This indicates that saturating the GPUs requires a prefill chunk with input length of $(4K-32)$.
However, the corresponding latency is $505ms$, far above the typical TBT SLO target ($<100ms$).

\autoref{fig:sweet_spot}-(b) presents TBT in Chunked-prefill with varying reused context lengths of the prefill. In this experiment, the token budget is fixed at 512, and the reused context length of decode iteration is 1K. As shown, TBT increases noticeably after the reused context exceeds $4K$. This reused context length is common in long-context understanding and multi-turn workloads, as shown in \autoref{tb:tasks}. In such cases, Chunked-prefill easily leads to SLO violations.

\subsection{New Paradigm \& Challenges}\label{sec:opportunity}
\begin{figure}
    \centering
    \includegraphics[width=.84\linewidth]{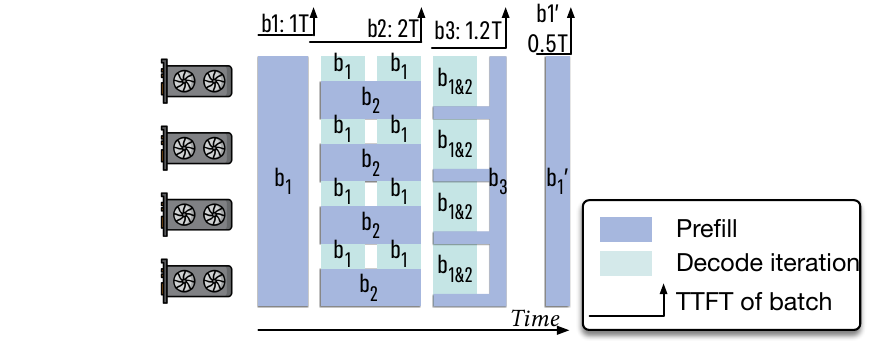}
    \caption{An ideal solution: prefill-decode multiplexing.}
    \label{fig:pd-multiplexing}
\end{figure}

As shown in \autoref{fig:pd-multiplexing},
we propose an intra-GPU prefill-decode (PD) multiplexing paradigm to overcome the above limitations.
Specifically, prefill and decode dynamically share the compute resources (SMs) within each GPU.
By reserving sufficient SMs to satisfy decode SLOs and assigning the remaining SMs to prefill,
high-goodput LLM serving is achieved.
PD multiplexing overcomes the limitations of prior methods, benefiting from the following abilities.

First, multiplexing enables dynamic and adaptive compute management. As shown, compute resources can be flexibly allocated between the two phases to maximize system goodput while guaranteeing SLOs.
Second, multiplexing decouples compute from memory management. Although the two phases partition compute resources, they share the memory space on each GPU, enabling efficient KV cache reuse.
Third, multiplexing allows prefill and decode to run independently without stalling one another, avoiding the dilemma between SLO attainment and system goodput.

However, integrating intra-GPU multiplexing into existing LLM serving systems is non-trivial. There are two challenges to realizing this paradigm. \textbf{C-1: GPU bubbles from naive integration.} Current systems have frequent prefill–decode interactions due to the inflight batching mechanism. One phase can easily block the other, creating GPU bubbles. \textbf{C-2: Unmanaged contention in spatial multiplexing.} Existing techniques~\cite{nvidia2025mig,nvidia_mps,nvidia2025grecontexts} partition only SMs while leaving memory bandwidth unmanaged. As a result, memory bandwidth contention can lead to SLO violations.

%% file: section/overview.tex
\section{\sysname{}'s Design}

\subsection{Architecture Overview}
\begin{figure}
    \centering
    \includegraphics[width=\linewidth]{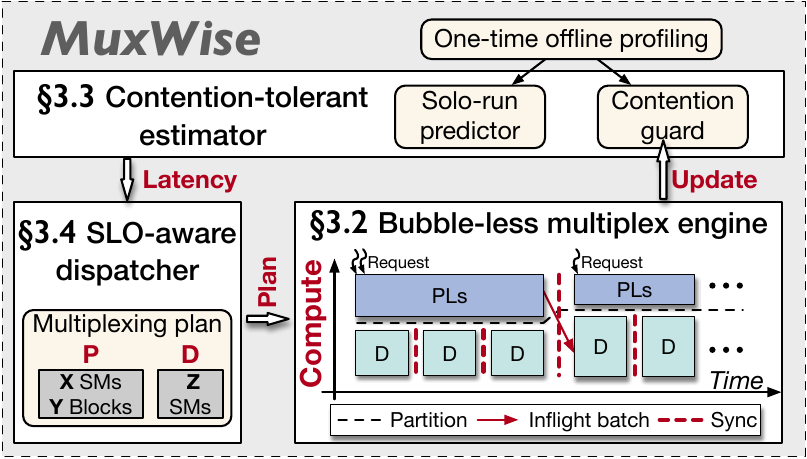}
    \caption{Architecture overview of \sysname{}.}
    \label{fig:overview}
\end{figure}
Based on the PD multiplexing paradigm, we propose \sysname{}, an LLM serving framework
that achieves high goodput on diverse workloads.
\autoref{fig:overview} shows the overview of \sysname{} that comprises (1) a bubble-less multiplex engine, (2) a contention-tolerant estimator, and (3) an SLO-aware dispatcher.

To enable PD multiplexing with bubble-less coordination, the engine partitions prefill into layer-wise execution, aligning its latency with decode execution.
Notably, layer-wise execution incurs negligible overhead and avoids the inefficiencies of chunk-prefill.
In addition, it provides an extra benefit: preempting ultra-long prefills to prevent the SLO violations caused by them.

To avoid SLO violations caused by unpredictable contention, the estimator provides worst-case latency estimates by combining a solo-run latency predictor with a contention guard.
Both components are built from one-time offline profiling for each LLM on a given hardware.
They consider five key factors: reused length, input length, output length, decoding batch size, and partition configuration.
LLM-specific factors are extracted from workload traces to guide profiling.

As requests arrive, the SLO-aware dispatcher leverages the engine and estimator to schedule prefill layers and decode iterations dynamically for high goodput.
Specifically, the dispatcher reserves best-fit SMs to satisfy decode SLOs based on the worst-case estimation, and assigns the remaining SMs to prefill.
Meantime, during online serving, it further refines the contention guard using runtime execution data.

%% file: section/pooling_compute.tex
\subsection{Bubble-less Multiplex Engine}\label{sec:gang_scheduling}

\subsubsection{Spatial Multiplexing Technique}

According to the analysis in \S\ref{sec:opportunity}, \sysname{} imposes two requirements for PD multiplexing. First, the compute resources must be dynamically partitioned between the two phases with low overhead. Second, the memory space must be shared across the phases to enable efficient KV cache reuse. To meet these requirements, we examine existing approaches for spatially partitioning GPU compute across tasks.

We categorize these approaches into two types: inter- and intra-process partitioning. Inter-process approaches, such as CUDA MIG~\cite{nvidia2025mig} and CUDA MPS~\cite{nvidia_mps}, cannot provide flexible compute resource adjustment, let alone the introduced cross-process communication between prefill and decode. In contrast, the intra-process approach GreenContext~\cite{nvidia2025grecontexts} enables low-overhead resource adjustment by binding CUDA streams to specific SMs, with reconfiguration costing only a stream synchronization (on the order of microseconds). Furthermore, because both phases reside in the same process under GreenContext, they can directly share the same memory space for maintaining a single KV cache pool.

\subsubsection{Inefficiencies from Naive Integration}
\begin{figure}
    \centering
    \includegraphics[width=\linewidth]{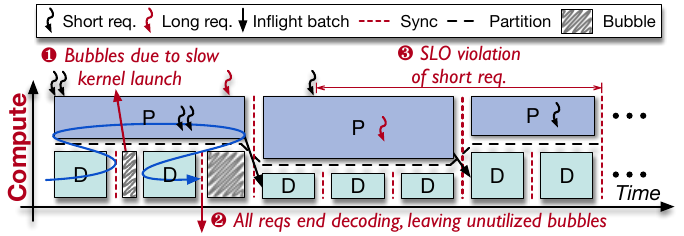}
    \caption{Bubbles and SLO violation of naive intra-process multiplexing. \textcolor{blue}{$\rightarrow$} represents the kernel launch order.}
    \label{fig:original_scheduling}
\end{figure}

With intra-process compute partitioning, a naive way to support PD multiplexing is to launch the ongoing decode iteration before the prefill phase of new request.
This ordering is motivated by launch latency difference: launching a decode iteration takes less than 0.5 ms, whereas launching a prefill phase takes tens of milliseconds.

Ideally, the launch latency of either a prefill phase or decode iteration can be reduced to a single CUDA graph launch (\textasciitilde{}$0.5 ms$).
However, CUDA graph requires offline construction with static configuration and incurs memory overhead.
In prefill phase, both batch size and input length vary, whereas decode iteration varies only in batch size.
Thus, single-graph optimization is feasible only for decode phase with several selected batch sizes~\cite{cudagraph}, while applying it to prefill phase would require capturing much more graphs, incurring unacceptable memory overhead.

Prefill phase can be optimized through piecewise CUDA graph~\cite{cudagraph}, which splits the prefill phase into multiple layer-wise CUDA graphs.
It still incurs \textasciitilde{}$\textbf{10 ms}$ launch overhead for Llama-70B on 8 A100 GPUs.
Fortunately, prefill phases consists of long-duration kernels, which are typically longer than the launch time.
It does not suffer noticeable performance degradation from launch overhead in most cases.

To this end, when both a prefill phase and a decode iteration are pending, \sysname{} prioritizes launching the decode iteration; otherwise, the SMs allocated for the decode phase would remain idle for tens of milliseconds.
However, this naive approach still introduces two types of GPU bubbles and can also lead to SLO violations.

Firstly, when the prefill launch time exceeds the execution time of a decode iteration, next decode iteration cannot launch in time, and GPU bubbles occur. As shown on the left of \autoref{fig:original_scheduling}, a bubble appears between two decode iterations because the serving system must return newly generated tokens after each iteration.

Secondly, bubbles can arise from unpredictable termination of decode. As depicted in the middle of \autoref{fig:original_scheduling}, all requests in a decode batch may finish token generation while a concurrent prefill phase has already been launched. Due to the non-preemptive nature of GPU execution, the launched prefill cannot be interrupted to reclaim compute resources.

Thirdly, SLO violations may occur due to workload skew among requests, as illustrated in the upper-right of \autoref{fig:original_scheduling}. Context lengths can vary significantly, with short conversations coexisting alongside long-text summarizations. In such cases, a short request may suffer long queuing delays while waiting for the prefill of an ultra-long request. If the short request has limited SLO slack, it is likely to miss its deadline.

\subsubsection{Bubble-less Coordination}

The above inefficiencies stem from the large latency discrepancy between the prefill and decode phases.
This is because prefill phases typically take longer to launch and execute, and the execution time of both phases is highly variable at runtime. To address this, we propose layer-wise execution for prefill and query-based synchronization.

\begin{figure}
    \centering
    \includegraphics[width=\linewidth]{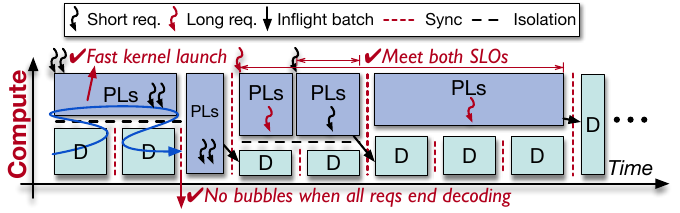}
    \caption{Bubble-less coordination using layer-wise scheduling for the prefill phase and graph-level scheduling for the decode phase. \texttt{PLs} is the short for prefill layers.
    }
    \label{fig:adaptive-gang}
\end{figure}

\mypar{Layer-wise execution for prefill}
As shown in \autoref{fig:adaptive-gang}, \sysname{} splits the prefill phase into layers (PLs). Based on this new granularity, \sysname{} eliminates GPU bubbles and prevents SLO violations. For the first type of bubble, \sysname{} can launch enough PBs to occupy compute resources for prefill, and return in time before the decode phase finishes computation. For the second type, \sysname{} switches the execution of later prefill layers into a new GreenContext, just after the decode phase terminates. For SLO violations caused by long requests, layer-wise execution enables preemption, allowing short requests to be prioritized, thereby meeting the SLO targets of both. Importantly, layer-wise execution incurs negligible overhead, since LLMs are inherently structured as multiple transformer layers.

\mypar{Query-based synchronization}
In addition to the above bubbles, inflight batching can also introduce GPU bubbles. Specifically, when the last prefill layer completes, it must block the next decode iteration to merge requests into the decode batch. This blocking creates small bubbles, since prefill and decode rarely finish simultaneously. To address this, \sysname{} employs query-based synchronization that periodically polls CUDA events. \sysname{} continues launching decode batches and prefill layers asynchronously, and when an event is observed complete, the corresponding prefill request is immediately merged into the current decode batch.

\subsection{Contention-tolerant Estimator}\label{sec:modeling}

When the prefill and decode phases are spatially multiplexed, contention can arise, particularly from unmanaged resources such as memory bandwidth. We begin by analyzing contention between the two phases under spatial multiplexing and then introduce our modeling method.

\begin{figure}
    \centering
    \includegraphics[width=\linewidth]{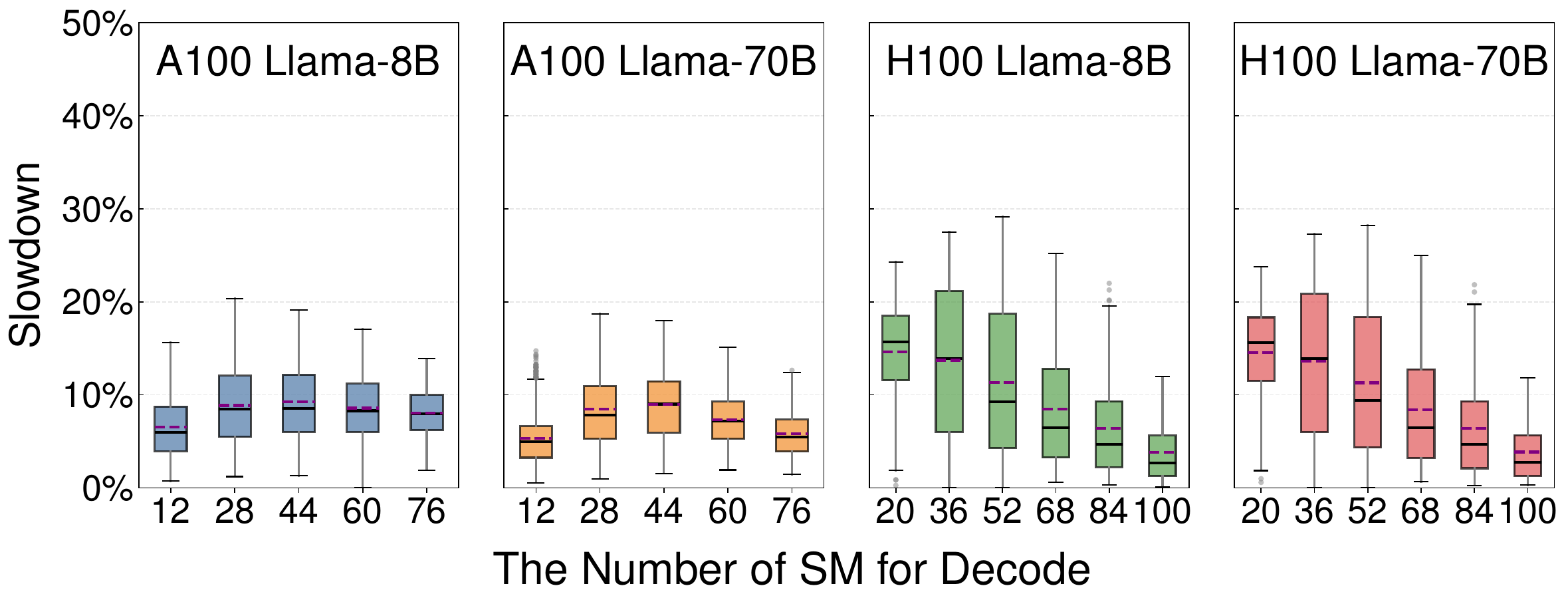}
    \caption{Slowdowns in decode due to contention with different multiplexing configurations of prefill and decode across models and GPUs.}
    \label{fig:contention-profiling}
\end{figure}

\subsubsection{Contention Analysis}

While GreenContext supports precise compute resource allocation, it cannot manage memory or network bandwidth.
In particular, efficient techniques for bandwidth management are lacking. Worse, current GPUs do not expose runtime monitoring of bandwidth usage, and both prefill and decode phases can heavily consume bandwidth, making contention hard to predict.

To evaluate the impact on execution slowdown, we extensively profile prefill and decode under multiplexing using Llama-8B and Llama-70B.
\autoref{fig:contention-profiling} reports the decode slowdown on servers with 8 A100 and 8 H100 GPUs. The x-axis denotes the number of SMs allocated to the decode batch, with the remaining SMs assigned to the prefill batch. For each configuration, the prefill batch’s total context length (reused + new) ranges from 1,024 to 128K tokens, whereas the decode batch’s reused length ranges from 1,024 to 1,024K tokens.
This profiling takes one week.

As shown in \autoref{fig:contention-profiling}, contention-induced slowdown ranges from nearly zero to about $30\%$ across different partition configurations and GPUs. The high variation across hardware partitions indicates the inherent unpredictability of contention slowdown. Although both models exhibit similar slowdown trends on the same GPUs due to their architectural similarity, this observation does not aid contention modeling. Meanwhile, results for prefill are similar but omitted due to space constraints.

\subsubsection{Worst-case Estimation for SLO guarantee}

To mitigate the risk of SLO violations caused by unpredictable contention in online serving, \sysname{} introduces a worst-case latency estimation method tailored for SLO guarantees. \textit{The key observation is that precise latency prediction is not the only way to guarantee SLO. What matters is ensuring that the latency of a scheduled phase, given its allocated compute resources, does not exceed the predefined target.} Thus, \sysname{} performs worst-case estimation by first predicting its solo-run latency, and then applying a maximum slowdown factor.

\mypar{Solo-run predictor}
To predict solo-run latency, we analyze the compute complexity of prefill and decode, and construct a predictor using offline profiling. The latency of each prefill or decode iteration is determined by the token lengths of the reused and new context. \autoref{tab:theory_cost} summarizes the compute analysis of the prefill and decode phases with a batch size of 1. The key factors are as follows:

\begin{itemize}
[leftmargin=*,topsep=0.em]
    \item $d$: The hidden dimension of each token’s representation.
    \item $L$: The total token length.
    \item $r$: The token length of the reused (cached) context.
    \item $n = L - r$: The token length of the new context.
\end{itemize}

\begin{table}
\centering
\footnotesize
\caption{Compute analysis for prefill and decode phases.}
\label{tab:theory_cost}
\begin{tabular}{c|cc}
\toprule
 & \textbf{Attention} & \textbf{FFN} \\
 \midrule
\textbf{Prefill w/o cache} & $\mathcal{O}(Ld^2+L^2d)$ & $\mathcal{O}(Ld^2)$ \\
\textbf{Prefill w/ cache} & $\mathcal{O}(nd^2 + Lnd)$ & $\mathcal{O}(nd^2)$ \\
\textbf{Decode} & $\mathcal{O}(d^2+(r+1)d)$ & $\mathcal{O}(d^2)$ \\
\bottomrule
\end{tabular}
\end{table}

Based on the complexity analysis in \autoref{tab:theory_cost}, we build latency prediction models for the prefill and decode phases. The prefill model is given in \autoref{eq:prefill}, and the decode model in \autoref{eq:decode}, where all $\theta$ terms are coefficients. We separate the models because state-of-the-art serving frameworks adopt different execution paths for these two phases, including distinct GPU kernel implementations and launch methods.

{
\footnotesize
\begin{align}
    \label{eq:prefill}
    T_{Prefill} &= \theta_1\cdot\sum_i^{bs}{n_{i}^2} + \theta_2\cdot\sum_i^{bs}{n_{i}\cdot r_{i}} + \theta_3\cdot\sum_i^{bs}{n_{i}} + \theta_4\\
    \label{eq:decode}
    T_{Decode} &= \theta_1\cdot\sum_i^{bs}{r_{i}} + \theta_2\cdot{bs} + \theta_3
\end{align}
}

The trained models achieve high accuracy, with a maximum deviation of $8.16\%$ for prefill and $8.84\%$ for decode, effectively supporting \sysname{}’s online scheduling.
Meanwhile, the offline profiling for training the solo-run predictor can be completed within a few hours, which is acceptable.
It is a one-time effort per LLM–machine pair and has been widely adopted in prior LLM serving work~\cite{wuLoongServeEfficiently,agrawalTamingThroughputlatency,zhongDistServeDisaggregating}.

\mypar{Contention guard}
To provide the maximum slowdown factor, \sysname{} introduces the contention guard.
Specifically, the contention guard provides slowdown factors only for decode, which will be explained in \S\ref{sec:priority}.
The contention guard is built using data collected through grid-sampling-based profiling.
This profiling spans five variables: the number of new and reused tokens in prefill, the batch size, and total reused tokens in decode, and the partition configuration.
For each pair of prefill and decode iterations to be estimated, the contention guard returns the maximum slowdown factor of the grid cell they fall into as the estimation result.

Building such a contention guard incurs much higher offline profiling overhead than the solo-run predictor, since pairwise profiling is needed to obtain slowdown data.
Fortunately, extensive profiling shows that slowdown remains within a limited range, with a maximum of 20\% on A100 GPUs and 30\% on H100 GPUs.
This indicates that even with coarse-grained profiling, worst-case latency inflation does not exceed 30\%.

To this end, we initialize the contention guard using coarse-grained grid sampling.
Specifically, we sample variables such as new and reused tokens in prefill, as well as single-request reused tokens in decode batches, at powers-of-4 granularity, ranging from 2K to 128K.
The sampled decode batch sizes follow SOTA serving frameworks (around 20 batch sizes).
We partition GPUs at the granularity of 16 SMs, yielding 6 configurations for A100 and 7 for H100. In total, the number of samples per LLM–machine pair is calculated as $(4 \times 4-1)\times 4 \times 20 \times 6 \approx 7K$, which can be collected within 12 hours.
In the equation, we exclude the case with 128K new and 128K reused tokens in prefill, since 128K is the maximum context window supported by mainstream LLMs~\cite{dubeyLlama3,baiQwenTechnical}.

The reason for using 16 SMs as the granularity are twofold:
(1) kernels on H100 and newer GPUs requires 16 SMs, due to using new features like thread block cluster~\cite{cuda-cluster}, and
(2) experiments indicate that 16 SMs already deliver strong performance improvements. Finer-grained scheduling offers little benefit while increasing memory overhead.

Furthermore, \sysname{} leverages runtime execution data to continuously update the contention guard, thus refining its SLO guarantees.
Even with the coarse-grained contention guard, \sysname{} already outperforms existing baselines in both SLO attainment and system goodput (\S\ref{sec:eval}).

\subsection{SLO-aware Dispatcher}\label{sec:dispatch}
With bubble-less multiplexing and contention-tolerant modeling, we introduce \sysname{}'s detailed dispatching policy.

\subsubsection{Priorities of prefill and decode}
\label{sec:priority}
In this work, we focus on the scheduling within a single serving instance.
In \sysname{}, we prioritize SLO attainment for the decode phase and process the prefill phase as early as possible.
SLO attainment for the prefill phase is not directly guaranteed for two reasons.
Firstly, although we prioritize the decode phase, we only allocate just-enough compute resources for it. Since the remaining compute resources are allocated to the prefill phase, its SLO is generally expected to be met.
Secondly, when SLO violations occur for the prefill phase, it indicates that the inference load has exceeded the peak capacity of the current LLM serving instance.
In such cases, further scheduling efforts would no longer improve performance.

This is also why the contention guard in the contention-tolerant estimator only provides a maximum slowdown factor for decode.
When predicting the prefill phase, \sysname{} does not need an accurate or worst-case estimate.
It only requires that the predicted latency of the launched prefill layers exceeds that of the corresponding decode iteration, ensuring full utilization of the allocated compute resources.

\subsubsection{Dispatching policy}
\begin{figure}
    \centering
    \includegraphics[width=\linewidth]{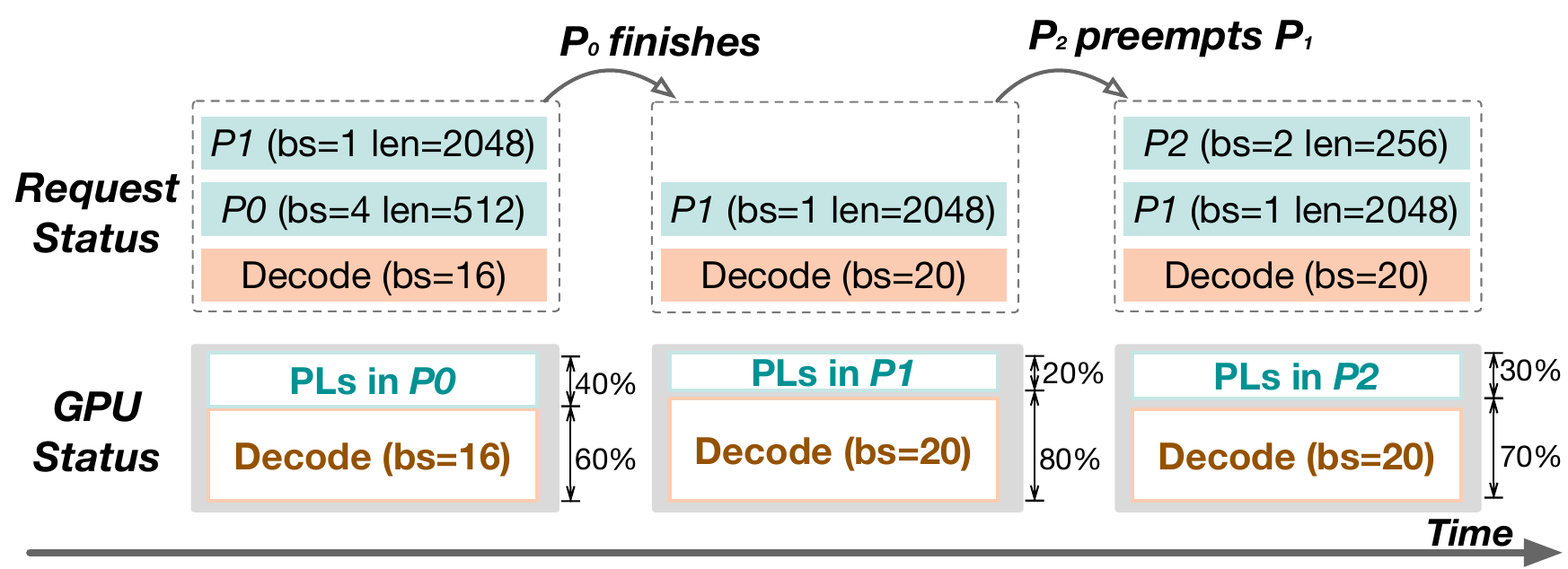}
    \caption{The dispatching policy of \sysname{}.}
    \label{fig:dispatcher}
\end{figure}

Building on the above analysis, \autoref{fig:dispatcher} illustrates \sysname{}’s SLO-aware dispatching policy. The system makes scheduling decisions after each prefill batch completes and at the end of each decode iteration. Specifically, the dispatcher selects prefill layers either from the ongoing prefill batch or from a new batch in the request queue, and allocates compute resources between the prefill and decode phases.

As shown in \autoref{fig:dispatcher}, the dispatcher allocates a best-fit number of SMs (60\%) to decode and assigns the remaining SMs (40\%) to prefill phase $P0$.
The resource partition satisfies the decode SLO and maximizes prefill throughput guided by the contention-tolerant estimator.
To support layer-wise execution in the multiplex engine, the estimator computes the number of prefill layers to launch as $N_{PL} = \lceil (T_d\times N_T) / T_P\rceil$, where $T_d$ is the estimated decode latency, $T_P$ is the estimated prefill latency, and $N_T$ is the number of transformer layers in the served LLM.

Once $P0$ finishes computation, it is merged into the decode batch, increasing the batch size to 20. The scheduler then retrieves a new prefill batch $P1$ and adjusts the partition to 20\% SMs for prefill and 80\% for decode to meet SLO targets.

Later, when a new prefill batch $P2$ arrives, it would normally wait for $P1$ to complete. However, because $P1$ has a long input length, this delay risks violating P2’s SLO. To avoid this, \sysname{} allows $P2$ to preempt $P1$, provided that preemption does not cause $P1$ to miss its own TTFT SLO.

\sysname{} does not allow recursive preemption. For example, after $P2$ preempts prefill batch $P1$, no other batch may preempt $P2$. This design is reasonable, as short requests typically preempt long ones, and preempting a short request in turn would likely cause it to miss its SLO. \sysname{} checks SLO attainment only when a prefill batch is preempted; otherwise, it prioritizes processing the active prefill batch as quickly as possible. Notably, preemption in \sysname{} is optional. Even when disabled, \sysname{} still delivers substantial performance improvements over the baselines.

%% file: section/evaluation.tex
\section{Evaluation}
\label{sec:eval}

\subsection{Experimental Setup}\label{eval:settings}

\mypar{Testbed}
We mainly evaluate \sysname{} on a server equipped with 8 A100-80GB GPUs.
The GPUs are interconnected via NVLINK, providing 600 GB/s of bandwidth.
We also evaluate \sysname{} on two additional servers to demonstrate its effectiveness on newer GPUs and larger LLMs: one with 8 H100-SMX5-80GB GPUs and another with 8 H200-SMX5-141GB GPUs.
These servers offer higher compute capability and larger GPU memory.
All experiments are conducted with PyTorch 2.6.0~\cite{paszkePyTorchImperative}. \sysname{} is implemented using SGLang~\cite{zhengEfficientlyProgramming} version 0.4.10post2. The GPU driver version is 570.124.06, and the CUDA version is 12.8.

\mypar{Models}
We primarily evaluate \sysname{} using two LLMs from the Llama family~\cite{touvronLLaMAOpen,touvronLlama2,dubeyLlama3}: Llama-8B and Llama-70B. These models differ in size and represent the most commonly hosted LLMs in the cloud.
We also evaluate a larger MoE model, Qwen3-235B with 22B activated, to demonstrate \sysname{}’s generality.

\mypar{Baselines}
We compare \sysname{} against 3 state-of-the-art solutions for efficient LLM serving.
Model parallelism techniques such as tensor parallelism~\cite{zhengAlpaAutomating} are employed to parallelize the deployed models.
For \sysname{}, we fix the tensor parallelism degree to 8.
Details of each baseline’s model parallelism configuration are provided when the baseline is introduced. For all systems, the KV cache memory pool is configured as large as possible to maximize throughput.
\begin{itemize}
[leftmargin=*,topsep=0.1em]
   \item \textbf{Chunked-prefill in SGLang~\cite{agrawalTamingThroughputlatency}}: This version of SGLang is equipped with chunked-prefill, as proposed by SARATHI-Serve~\cite{agrawalTamingThroughputlatency}.
    We follow SARATHI-Serve’s methodology to calculate the token budget for each workload prior to experiments.
    It is offline tuned under specific TBT targets for each model.
    Unlike SARATHI-Serve, which serially executes prefill and decode attention kernels, SGLang leverages Flashinfer~\cite{yeFlashInferEfficient}, a high-performance inference kernel library, to fuse them into a single kernel.
    It is expected to deliver performance similar to POD-attention~\cite{kamath_pod-attention_2025}.

    \item \textbf{NanoFlow~\cite{zhu_nanoflow_2025}}: This is an enhanced version of chunked-prefill with operator-level intra-GPU multiplexing, targeting near-optimal throughput under a relatively loose SLO requirement ($200$ ms). It requires a large token budget (at least 1024) to achieve this goal. However, such a token budget cannot meet the SLO requirements of modern LLM serving ($\leq100$ ms).
    We use the same token budget as chunked-prefill for NanoFlow.
    
    \item\textbf{LoongServe~\cite{wuLoongServeEfficiently}}: This is a dynamic disaggregated serving system.
    We adopt its model-parallelism configuration.
    For Llama-70B, sequence parallelism is set to 2 and tensor parallelism to 4.
    For Llama-8B, sequence parallelism is set to 4 and tensor parallelism to 2.
    It does not support new LLMs like MoE models.
    \item \textbf{Disaggregated serving in SGLang (SGLang-PD in short)}:
    This is the latest implementation of static disaggregation with KV-cache sharing across phases and requests.
    The P:D ratio is 1:1, with tensor parallelism set to 4 for each instance.
    DistServe~\cite{zhongDistServeDisaggregating} does not support KV-cache sharing across requests, making it unsuitable for modern LLM services.
    We also evaluated Dynamo~\cite{nvidia2025dynamo}, which performed substantially worse than SGLang-PD.
    Therefore, we use SGLang-PD as the state-of-the-art baseline of static disaggregation for evaluation.
\end{itemize}

\mypar{Metrics}
\sysname{} targets goodput improvement. So, following prior works~\cite{zhongDistServeDisaggregating, agrawalTamingThroughputlatency, patelSplitwiseEfficient,qinMooncakeTrading}, we use tail latency (e.g., P99) to assess SLO attainment.
Meanwhile, there is also another metric to measure the SLO guarantee during decode phase: TPOT (timer per output token).
In comparison, TBT accounts the latency of each individual token, whereas TPOT is an average metric that may mask the poor performance of some tokens~\cite{llmSlo}.
Thus, we choose TBT over TPOT for a stricter SLO metric.
We set the TBT SLO target to $50ms$ for Llama3-8B and $100ms$ for Llama3-70B, following prior works~\cite{agrawalTamingThroughputlatency,qinMooncakeTrading}.
We regard \sysname{}’s ability to deliver better TTFTs under skewed workloads as an additional benefit of the new serving paradigm.
Moreover, it breaks the first-come-first-serve model used in other baselines.
Thus, we only evaluate TTFT per token in \S\ref{eval:preempt}.

\begin{figure}
    \centering
    \includegraphics[width=\linewidth]{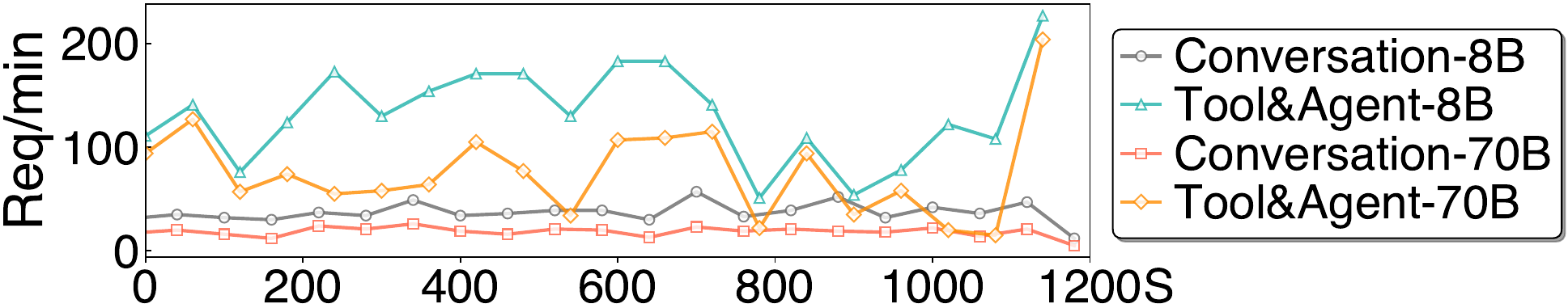}
    \caption{The two real-world workload traces after scaling.}
    \label{fig:analyze-trace}
\end{figure}

\begin{figure}
    \centering
    \includegraphics[width=\linewidth]{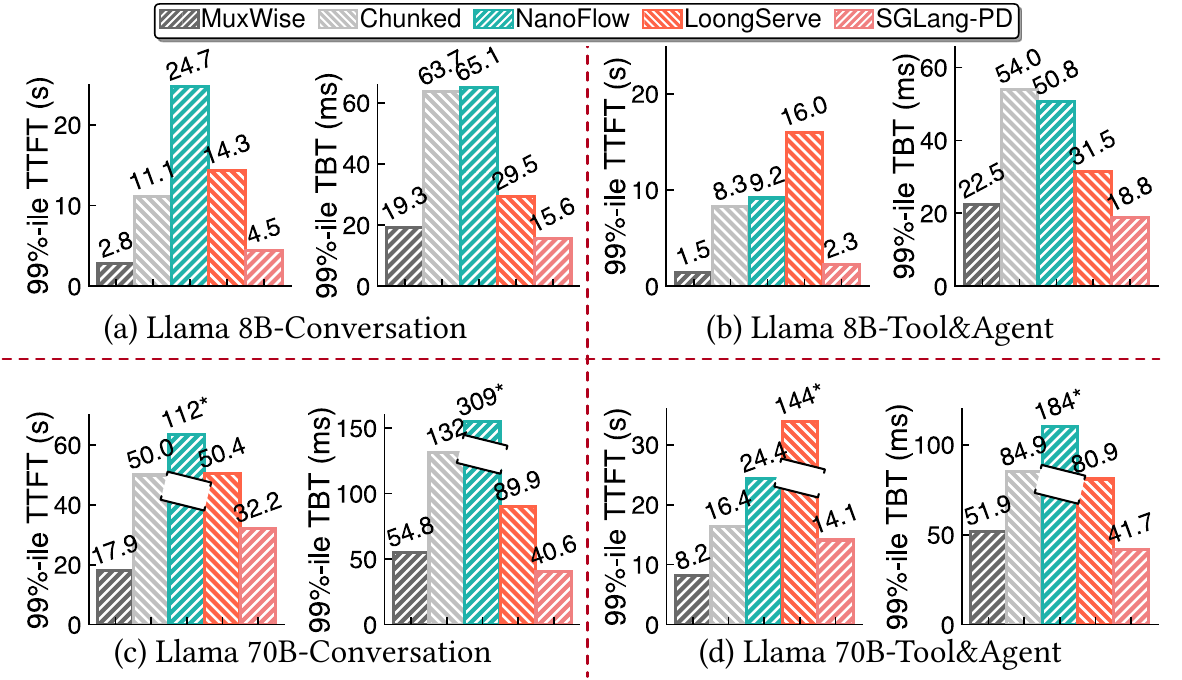}
    \caption{
    $99\%${\it -ile} TTFT and TBT for Llama-8B and Llama-70B on real-world Conversation and Tool\&Agent workloads. \texttt{Chunked} represents chunked-prefill in SGLang. Values marked with $*$ are too large; we clip their corresponding bars, so the bar height only decodes their relative size.
    }
    \label{fig:e2e-cdf}
\end{figure}

\subsection{Evaluation on Real-world Workloads}
\subsubsection{End-to-end Performance}\label{eval:real-world}
We begin by evaluating \sysname{} with Llama-8B and Llama-70B under real-world workload traces.
\autoref{fig:analyze-trace} shows their request rates after scaling down, as they are originally  from a large cluster.
As illustrated, they show bursty request patterns (up to $13\times$ spike within 1min).

\autoref{fig:e2e-cdf} shows the latency distribution of TTFT and TBT. Although the real-world traces are scaled down to a modest level,
some baselines still easily reaches its peak throughput and enters an unstable state (NanoFlow in \autoref{fig:e2e-cdf}-(c) and LoongServe in \autoref{fig:e2e-cdf}-(d)).
After omitting unstable results, \sysname{} achieves average 99\%{\it -ile} TTFT speedups of $3.57\times$, $5.98\times$, $4.65\times$, and $1.66\times$ over chunked-prefill, NanoFlow, LoongServe, and SGLang-PD, respectively.
\sysname{} and the two disaggregated solutions consistently meet the TBT SLO, whereas chunked-prefill and NanoFlow fails in most cases.
SGLang-PD achieves shorter TBT than \sysname{}, as it statically reserves more compute resources for the decode instance.

Compared to chunked-prefill, \sysname{} avoids the dilemma between SLO compliance and high utilization, bringing better performance for both prefill and decode phases.
While tuning the token budget, we observe that either increasing or reducing it fails for SLO guarantee.
This is because the reused length in prefill phase in the two workloads can reach up to 50K tokens.
Further splitting the prefill into smaller chunks does not help control the TBT.
\sysname{}’s PD multiplexing avoids this issue entirely.

NanoFlow performs worse than the original chunked-prefill.
This is because, built atop chunked-prefill, NanoFlow is designed to overlap compute-bound kernels with memory-bound or communication kernels.
To achieve this, it requires a large token budget (1024 in its paper) to ensure that chunked-prefill as a whole remains compute-bound.
However, in \autoref{fig:e2e-cdf}, the token budget has to be reduced to 256 to meet TBT SLO targets, where chunked-prefill is no longer compute-bound.
The long reused context length in the two evaluated real-world traces further makes chunked-prefill harder to be compute-bound.
Thus, NanoFlow degrades due to overlapping memory-bound kernels.

The situation worsen for NanoFLow with Llama-70B in \autoref{fig:e2e-cdf}-(c\&d).
This could be attributed to the inherent model weight reload of intra-GPU overlapping.
NanoFlow split each chunk into 2 nano batches , thus duplicating loading for each decode iteration~\cite{zhu_nanoflow_2025}.
When evaluating with Llama-70B, the reloading overhead is amplified due to the larger model size.
Conversely, \sysname{} duplicates loading only once during prefill, which typically co-runs with tens of decode iterations.
Because prefill is compute-intensive and the reload is amortized over the entire phase, \sysname{} imposes negligible bandwidth pressure, which is marginal relative to its overall benefits.
While NanoFlow performs poorly on the two real-world workloads, it outperforms chunked-prefill for short input sequences without cross-request context length reuse(\S\ref{eval:synthetic}).

Against the two disaggregated solutions, \sysname{} achieves significantly better TTFT.
In LoongServe, instance scaling releases the KV cache needed for reuse in the prefill phase, causing redundant recomputation.
In SGLang-PD, static disaggregation often leaves decode instances idle under fluctuating real-world workloads.
In contrast, \sysname{} avoids KV migration and adapts to dynamic workloads through intra-GPU compute partition reconfiguration.

\begin{table}
\setlength{\tabcolsep}{2pt}
\centering
\footnotesize
\caption{
Results of other metrics for Llama-70B on Conversation workloads in \autoref{fig:e2e-cdf}-(c).
}
\label{tab:other-metrics-conv}
\begin{tabular}{c|cccccccc}
\toprule
\multirow{2}{*}{} & \multicolumn{2}{c|}{TTFT (s)} & \multicolumn{2}{c|}{TBT (ms)} & \multicolumn{2}{c|}{E2E (s)} & \multicolumn{2}{c}{TPOT (ms)} \\ \cmidrule{2-9}
 & \multicolumn{1}{c|}{Avg.} & \multicolumn{1}{c|}{P50} & \multicolumn{1}{c|}{Avg.} & \multicolumn{1}{c|}{TBT} & \multicolumn{1}{c|}{Avg.} & \multicolumn{1}{c|}{P50} & \multicolumn{1}{c|}{Avg.} & \multicolumn{1}{c}{P50} \\
 \midrule

\sysname{} & \textbf{3.1} & \textbf{1.4} & \textbf{30.2} & \textbf{25.0} & \textbf{13.1} & \textbf{12.2} & \textbf{31.1} & \textbf{28.3} \\

Chunked & 12.0 & 7.2 & 45.3 & 49.3 & 27.0 & 23.3 & 46.9 & 45.7 \\

NanoFlow & 51.4 & 52.3 & 105.6 & 98.9 & 83.4 & 84.2 & 120.2 & 103.2 \\

LoongServe & 17.7 & 14.7 & 61.0 & 58.3 & 38.4 & 35.8 & 62.3 & 60.6 \\

SGLang-PD & 7.38 & 3.95 & 32.9 & 32.5 & 18.4 & 16.4 & 33.5 & 33.2 \\
\bottomrule
\end{tabular}
\end{table}

\begin{table}
\setlength{\tabcolsep}{2pt}
\centering
\footnotesize
\caption{Results of other metrics for Llama-70B on Tool\&Agent workloads in \autoref{fig:e2e-cdf}-(d).}
\label{tab:other-metrics-tool}
\begin{tabular}{c|cccccccc}
\toprule
\multirow{2}{*}{} & \multicolumn{2}{c|}{TTFT (s)} & \multicolumn{2}{c|}{TBT (ms)} & \multicolumn{2}{c|}{E2E (s)} & \multicolumn{2}{c}{TPOT (ms)} \\ \cmidrule{2-9}
 & \multicolumn{1}{c|}{Avg.} & \multicolumn{1}{c|}{P50} & \multicolumn{1}{c|}{Avg.} & \multicolumn{1}{c|}{P50} & \multicolumn{1}{c|}{Avg.} & \multicolumn{1}{c|}{P50} & \multicolumn{1}{c|}{Avg.} & \multicolumn{1}{c}{P50} \\
 \midrule

\sysname{} & \textbf{1.3} & \textbf{1.0} & \textbf{27.2} & 24.1 & \textbf{4.0} & \textbf{1.6} & \textbf{33.1} & \textbf{27.7} \\

Chunked & 2.4 & 1.1 & 30.5 & \textbf{21.2} & 5.5 & 2.3 & 45.9 & 47.1 \\

NanoFlow & 2.8 & 1.2 & 58.8 & 42.4 & 7.4 & 2.5 & 70.3 & 62.6 \\

LoongServe & 59.9 & 56.0 & 52.4 & 50.8 & 65.2 & 61.0 & 56.2 & 54.6 \\

SGLang-PD & 2.1 & 1.5 & 31.6 & 31.0 & 5.2 & 2.3 & 37.1 & 33.7 \\
\bottomrule
\end{tabular}
\end{table}

\subsubsection{Other Latency Metrics}
In this experiment, we report results for other metrics, such as end-to-end latency and TPOT.
We also present these results using other statistical measures, including average and P50 values.
\autoref{tab:other-metrics-conv} shows the results on the Conversation workload with Llama-70B, while \autoref{tab:other-metrics-tool} shows the results on the Tool\&Agent workload with Llama-70B in \S\ref{eval:real-world}.
Results in other settings are similar and are omitted due to space constraints.

As shown in the two tables, while \sysname{} focuses on improving high goodput, it also consistently outperforms the baselines across the reported metrics.
There is only one outlier in the P50 TBT of \autoref{tab:other-metrics-tool}, and the values are very close.
This can occur because the P50 TBT in chunked-prefill may correspond to the latency of a pure decode iteration.

\subsubsection{SLO Attainment and Goodput}\label{eval:slo}

We also measure the SLO attainment of TBT and the corresponding goodput to evaluate the effectiveness of \sysname{} in meeting SLO compliance.
In this experiment, we extract requests from the \texttt{Tool\&Agent} trace but replace their arrival timestamps with those generated by a Poisson process at varying rates, following prior work~\cite{wuLoongServeEfficiently}.
We stop testing once the serving system becomes unstable or fails to meet the TBT SLO target.

\begin{figure}
    \centering
    \includegraphics[width=\linewidth]{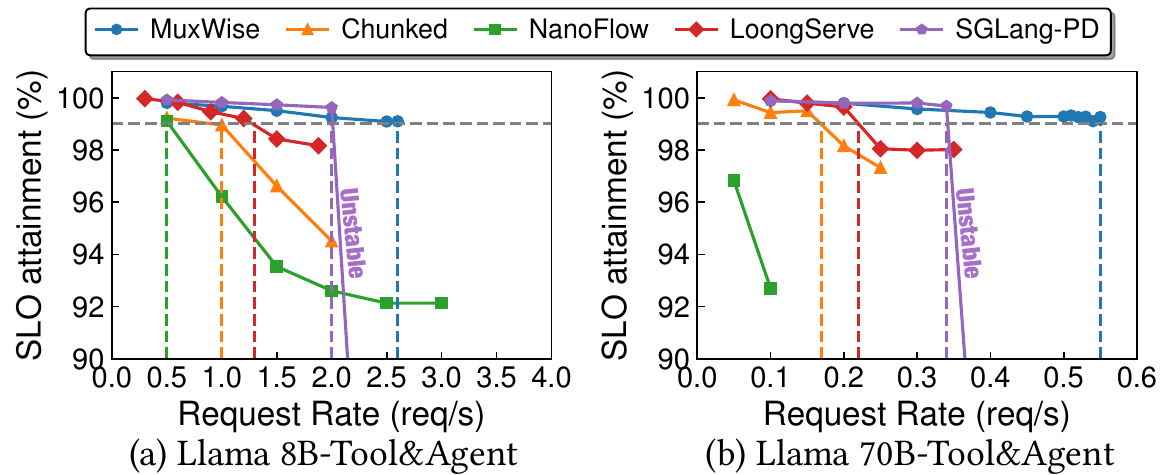}
    \caption{SLO attainment for Llama-8B and Llama-70B on Tool\&Agent workload with varied request rates.}
    \label{fig:slo-attainment}
\end{figure}

\autoref{fig:slo-attainment} shows the SLO attainment results under gradually increasing workloads.
Under the constraint of meeting the {\it 99\%-ile} SLO guarantees, \sysname{} achieves $2.6\times$, $5.2\times$, $2.0\times$, and $1.3\times$ higher goodput than chunked-prefill, NanoFlow, LoongServe, and SGLang-PD, respectively for Llama-8B; and $3.06\times$, $2.62\times$, and $1.62\times$ higher than chunked-prefill, LoongServe, and SGLang-PD for Llama-70B.
NanoFlow never meets the SLO even with a small chunk size of 64 for Llama-70B; therefore, the corresponding goodput improvement is omitted.
\autoref{tab:token-utilization} further shows the corresponding token thorughput and GPU utilization of \sysname{} and baselines.
GPU utilization is an aggregated metric reported by NVIDIA Nsight Systems, that reflects the fraction of active SMs as well as the utilization of intra-SM resources.

\begin{table}
\setlength{\tabcolsep}{2pt}
\centering
\footnotesize
\caption{Token throughput and GPU utilization for Llama-8B and Llama-70B on Tool\&Agent workload under Goodput.}
\label{tab:token-utilization}
\begin{tabular}{c|cccc}
\toprule
Model & \multicolumn{2}{c}{Llama-8B} & \multicolumn{2}{c}{Llama-70B} \\
\midrule
Metrics & \multicolumn{1}{c}{Token/s} & \multicolumn{1}{c}{GPU Util.} & \multicolumn{1}{c}{Token/s} & GPU Util. \\
\midrule
\sysname{} & 25397 & 88.1 & 7430 & 84.0 \\

Chunked & 9768 & 63.8 & 2269 & 66.1 \\

NanoFlow & 4884 & 55.1 & -- & -- \\

LoogServe & 12698 & 75.3 & 2936 & 70.1 \\

SGLang-PD & 19535 & P(72.4)/D(83.4) & 4538 & P(67.1)/D(81.9)\\
\bottomrule
\end{tabular}
\end{table}

Chunked-prefill, and NanoFlow fails to meet the TBT SLO even at lower request rates than the other two baselines.
This is because chunking is largely ineffective at reducing TBT in real-world LLM services, where cross-request interactions are common.
Compared to LoongServe, \sysname{} achieves higher goodput by avoiding recomputation in multi-turn requests.
Compared to SGLang-PD, \sysname{} achieves higher goodput through a larger KV-cache pool and reduced idleness caused by static disaggregation.
Meanwhile, \sysname{} achieves shorter TTFT across all cases (up to $9.16\times$).

\subsubsection{More Advanced GPUs and Larger LLM}

\begin{figure}
    \centering
    \includegraphics[width=\linewidth]{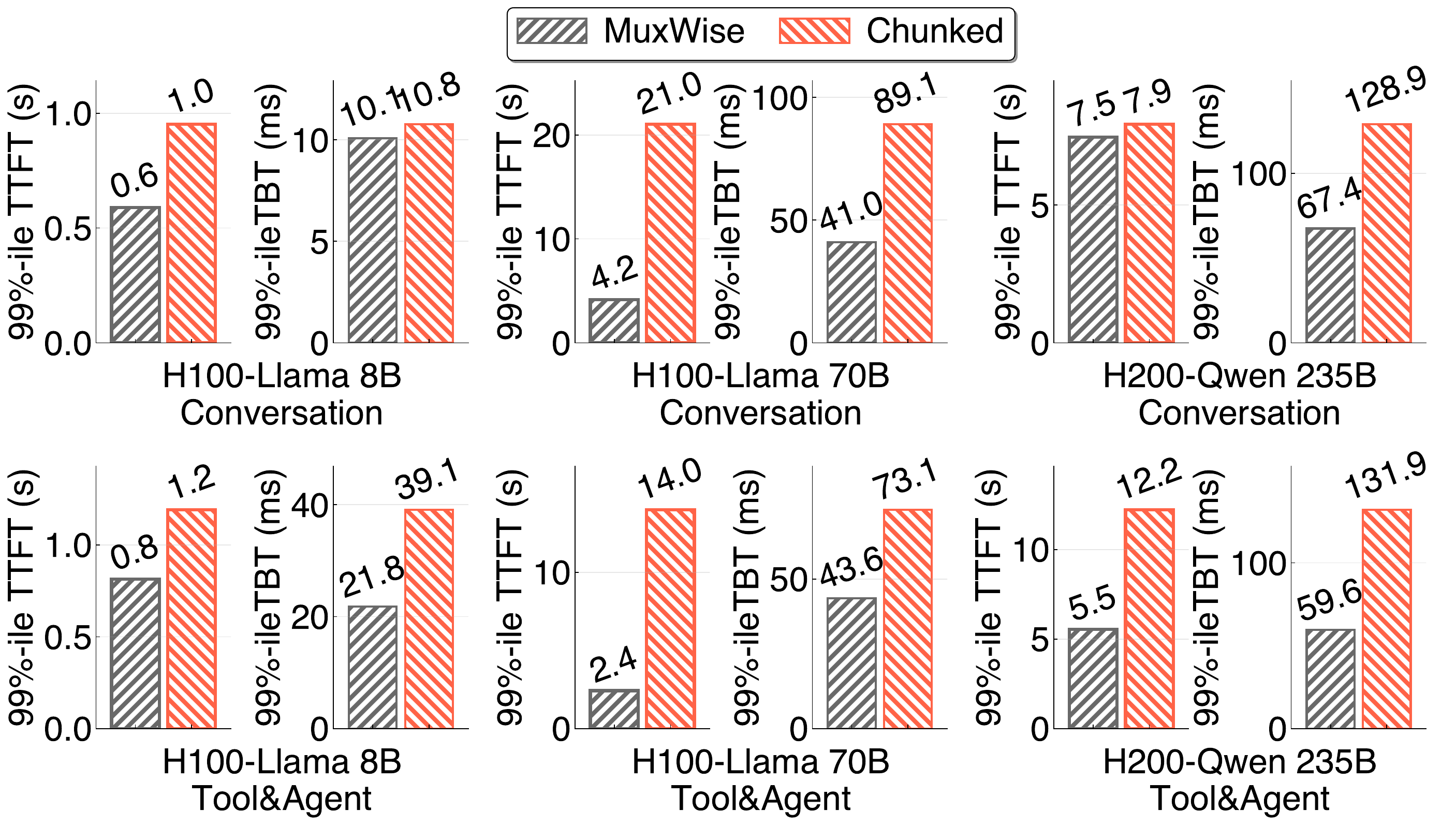}
    \caption{$99\%${\it -ile} TTFT and TBT for Llama-8B and Llama-70B on a server with 8 H100 GPUs and $99\%${\it -ile} TTFT and TBT for Qwen-235B on a server with 8 H200 GPUs.}
    \label{fig:other-gpus-llms}
\end{figure}

To demonstrate \sysname{}’s effectiveness on other GPUs and LLMs, we evaluate it with Llama-8B and Llama-70B on a server with 8 H100 GPUs, and with Qwen-235B on a server with H200 GPUs.
In this experiment, we only compare \sysname{} with chunked prefill.
LoongServe does not support new MoE models like Qwen-235B, and disaggregated serving solutions are also infeasible for Qwen-235B, even though each H200 has 141 GB of GPU memory.
\autoref{fig:other-gpus-llms} shows the experimental results.
Across all cases, \sysname{} achieves an average $2.28\times$ speedup on $99\%${\it -ile} TTFT and an average $1.81\times$ speedup on $99\%${\it -ile} TBT.
These consistent improvements demonstrate the generality of \sysname{}’s serving paradigm across diverse hardware and larger, newer LLMs.

\subsection{Evaluation on Diverse Synthetic Workloads}
\label{eval:synthetic}
To better demonstrate \sysname{}’s effectiveness, we further evaluate it under three synthetic workloads.
In the rest of the evaluation, we focus on Llama-70B due to space constraints, as results on other models are similar.
Requests are generated by sampling inputs from ShareGPT~\cite{sharegpt_2023}, Openthoughts~\cite{guha_openthoughts_2025}, and LooGLE~\cite{li2024looglelongcontextlanguagemodels}, with arrival rates gradually increased following a Poisson process.
Among these, only Openthoughts requests share a short system prompt.
We select these workloads because they represent three typical patterns: moderate input and output, short input with ultra-long output, and ultra-long input with short output.

\begin{figure}
    \centering
    \includegraphics[width=\linewidth]{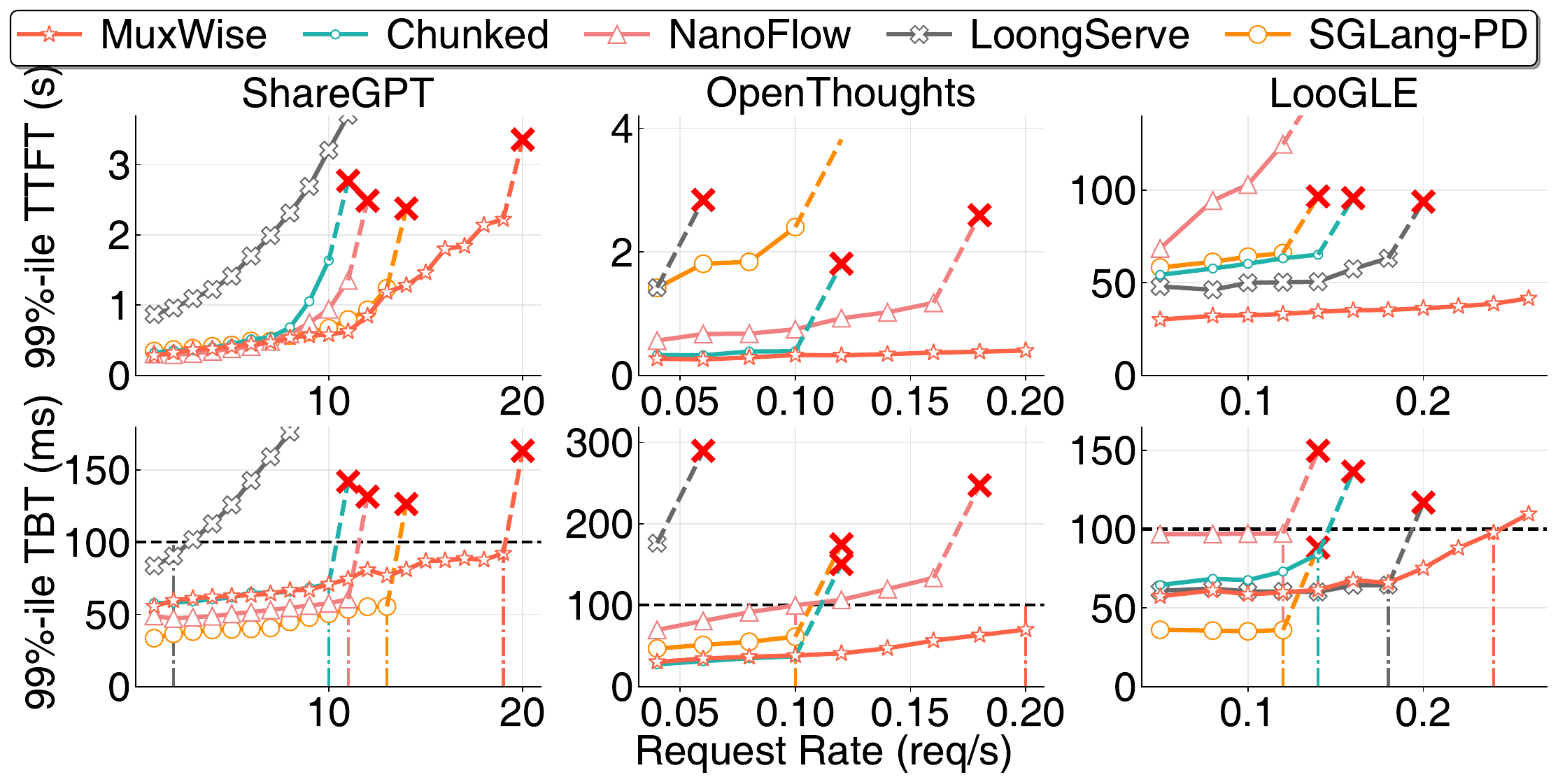}
    \caption{$99\%${\it -ile} TTFT and TBT with Llama-70B on three types of synthetic workloads.}
    \label{fig:e2e-synthetic}
\end{figure}

\autoref{fig:e2e-synthetic} shows {\it 99\%-ile} TTFT and TBT of \sysname{} and three baselines.
On ShareGPT, \sysname{} achieves goodput improvements of $1.9\times$, $1.73\times$, $9.5\times$, $1.46\times$ over chunked-prefill, NanoFlow, LoongServe, and SGLang-PD, respectively. On LooGLE, it achieves $1.71\times$, $2\times$, $1.33\times$, $2\times$ over the four baselines. On Open-Thoughts, it achieves the same $2\times$ improvement over chunked-prefill, NanoFlow and SGLang-PD, while Loongserve never meets SLO.

On ShareGPT, \sysname{}, chunked-prefill, NanoFlow, and SGLang-PD all provide SLO guarantees at the beginning.
SGLang-PD even achieves better TBT than \sysname{}, as it statically reserves more compute.
In contrast, \sysname{} delivers shorter TTFT by reserving only best-fit SMs for decode.
On OpenThoughts, LoongServe performs worse than the others, as it is designed for long-context workloads rather than requests with short inputs and long outputs.

NanoFlow outperforms chunked-prefill only on ShareGPT.
On OpenThoughts, the system spends most of the time in the decode phase.
Therefore, NanoFlow splits decode iterations to enable overlapping, leading to higher TBT than chunked-prefill.
On LooGLE, it performs worse due to the small token budget used for long requests.

We also observe that SGLang-PD performs much worse on OpenThoughts and LooGLE than on ShareGPT.
The causes differ across workloads.
For OpenThoughts, since requests share little context, the system must still reserve slots for KV caches during prefill and decode.
As the request rate increases, prefill stalls once the KV cache pool runs out of space.
For LooGLE, only four GPUs are available for prefill, causing requests to queue in the prefill instance.

\subsubsection{Short Requests and Single GPU}
Running Llama-8B on an A100 with ShareGPT, \sysname{} improves goodput by $1.2\times$ over chunked-prefill while maintaining similar TBT.
This is because even when chunking rarely happens, satisfying a strict TBT SLO still forces chunked-prefill to use a small token budget, limiting GPU utilization and peak goodput.
Notably, real-world conversation inputs are becoming significantly longer (e.g., 1.2K and 2.3K tokens in two recent conversation traces from cloud vendors~\cite{wang_kvcache_2025,stojkovic_dynamollm_2025}, compared with 226 tokens in the older ShareGPT dataset).
The conversation in our evaluation is a multi-turn real-world trace, whose average length approaches to 7.5K.
This trend is driven by the widespread adoption of techniques such as RAG~\cite{rag}, which increase the effective model input length by appending retrieved sequences to the user input.

\subsection{Ablation Study}
\subsubsection{Scheduling details of different tasks.}
\begin{figure}
    \centering
    \includegraphics[width=\linewidth]{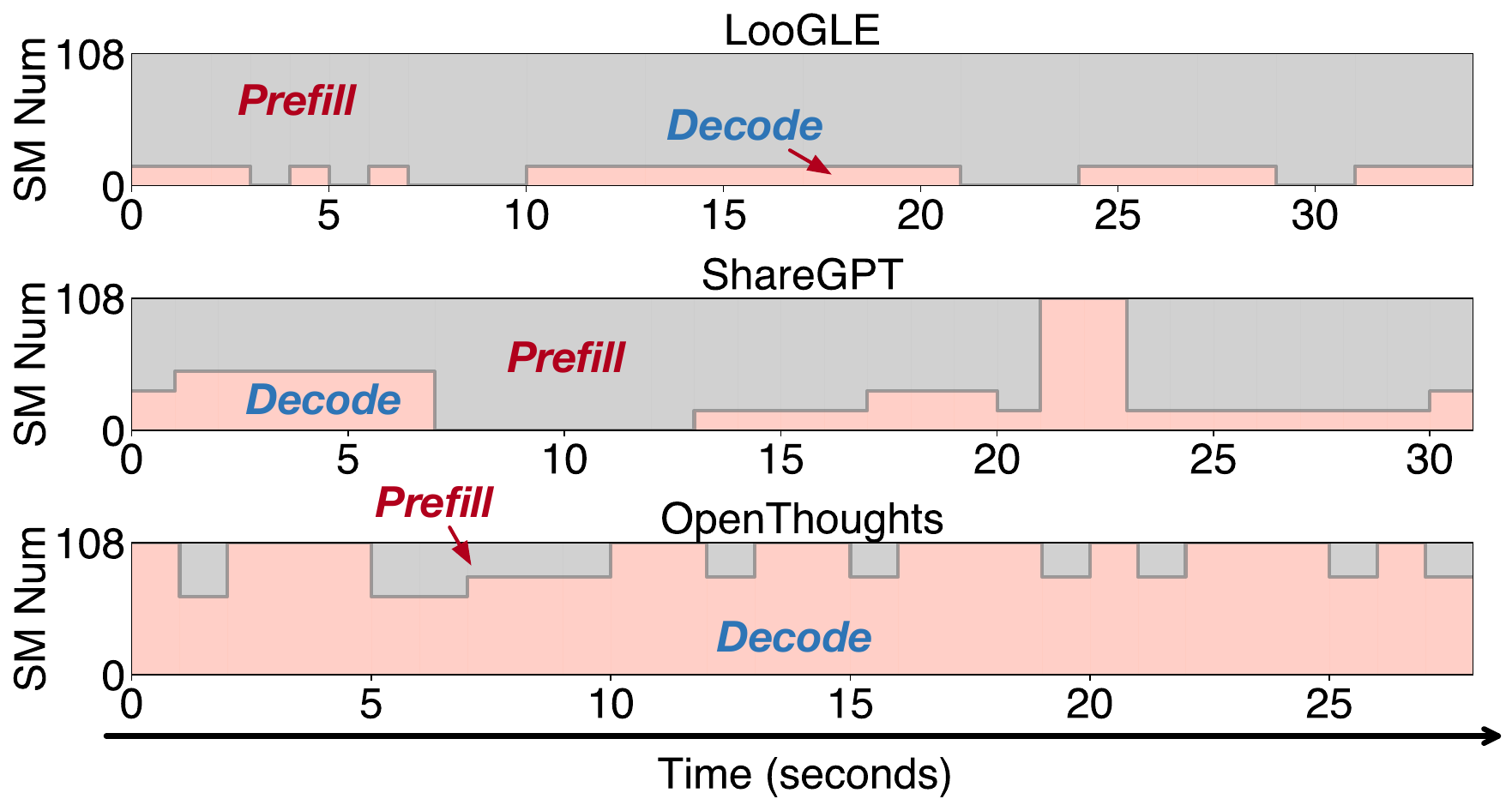}
    \caption{Change in compute partition between prefill and decode on LooGLE, ShareGPT, and OpenThoughts.
    Figures are sorted in descending order of prefill compute demand.
    }
    \label{fig:X-profiling}
\end{figure}

We further evaluate \sysname{}’s dynamic scheduling of compute partitions by extracting scheduling details from runtime serving in \S\ref{eval:synthetic}.
As shown in \autoref{fig:X-profiling}, \sysname{} makes different scheduling decisions for different workloads.
On LooGLE, most SMs are allocated to prefill, while on OpenThoughts, \sysname{} allocates the majority of SMs to decode.
Results on ShareGPT lie between LooGLE and OpenThoughts.
Overall, however, more SMs are allocated to prefill on ShareGPT, since decode is typically memory-bound and does not require as many SMs as prefill.
Notably, we use \autoref{fig:X-profiling} to show that different partitions are required for different workloads.
They are relatively static because the request rate is stable.
In real-world traces, the workload could be bursty.
Experimental results show that, during a bursty interval in \autoref{fig:analyze-trace}, \sysname{} activated all the six configurations within 30s.

\subsubsection{Effectiveness of Bubble-less Multiplex Engine}\label{eval:gang}
As shown in \autoref{fig:original_scheduling}, bubbles commonly occur in the green context created for decode.
In this experiment, we compare the TBT of \sysname{} against its two variants.
First, we disable layer-wise scheduling. Second, we further disable the query-based synchronization optimization.
The workloads used are Tool\&Agent under two different request rates.

\begin{figure}
    \centering
    \begin{minipage}[t]{0.48\linewidth}
    \centering
   \includegraphics[width=\linewidth]{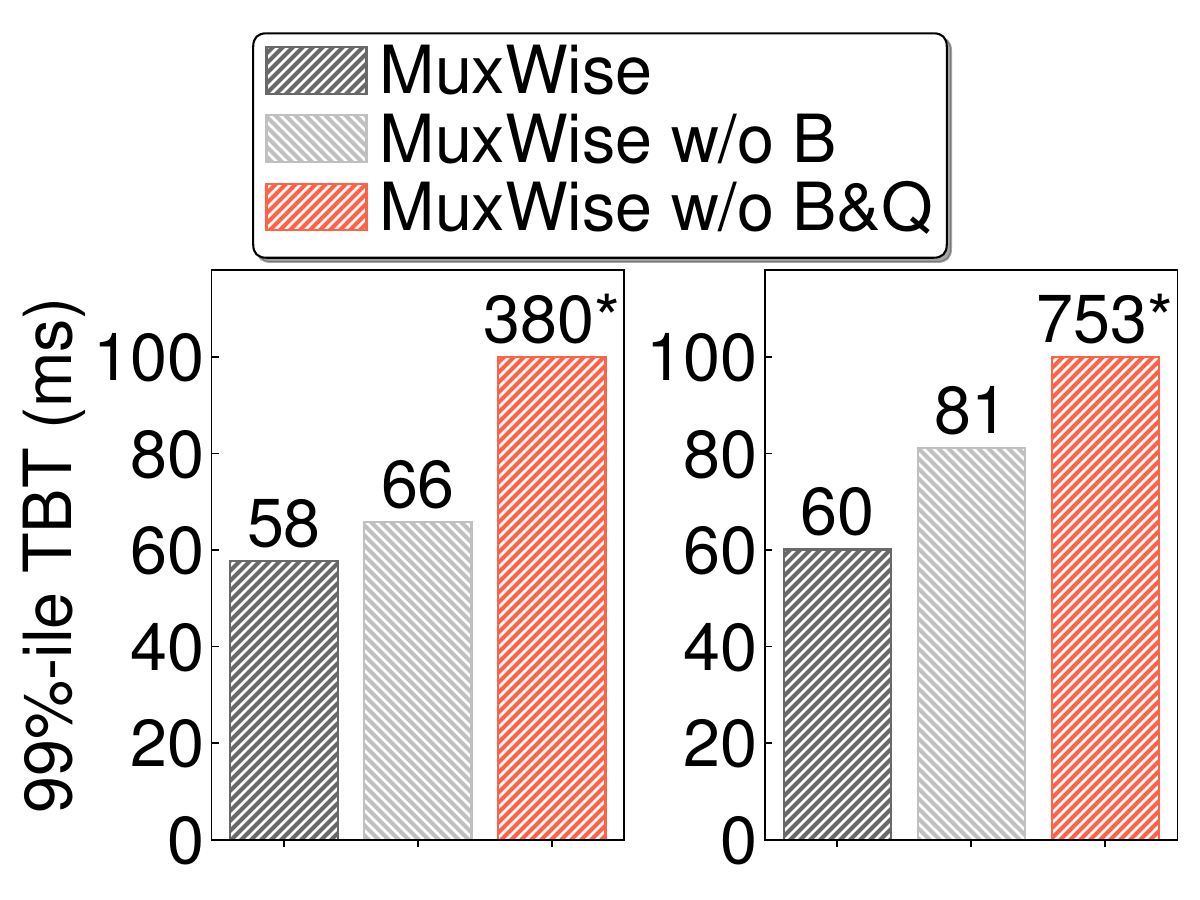}
    \captionof{figure}{\sysname{} with and without bubble-less multiplexing.}
    \label{fig:ablation-gang}
\end{minipage}
\hspace{1mm}
\begin{minipage}[t]{0.48\linewidth}
    \centering
   \includegraphics[width=\linewidth]{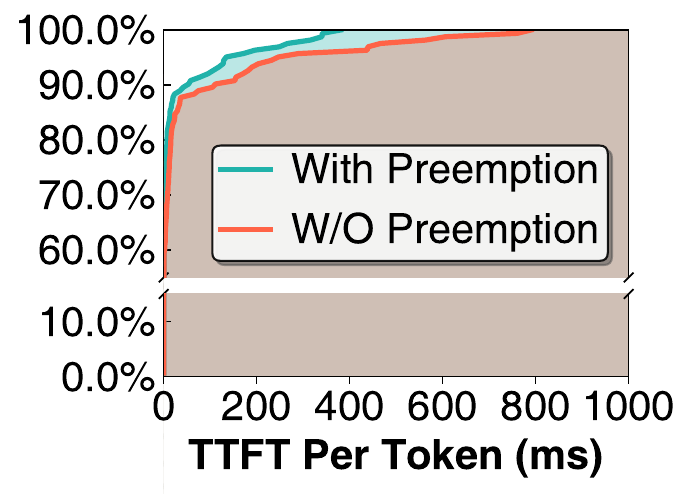}
    \captionof{figure}{CDF of TTFT per token with and without preemption.}
    \label{fig:ttft-per-token-cdf}
\end{minipage}
\end{figure}

\autoref{fig:ablation-gang} presents the experimental results.
As shown in the figure, disabling layer-wise execution slightly increases the TTFT of decode by approximately $10$ms, which aligns with the typical kernel launch time for the prefill phase of Llama-70B.
When query-based synchronization is further disabled, \sysname{} suffers a significant degradation, $314$ms for Llama-8B and $672$ms for Llama-70B, due to frequent stalls waiting for the prefill phase to complete.

For further evaluation, we also collect the bubble ratio of \sysname{} and chunked-prefill for the goodput results in \autoref{fig:slo-attainment}-a by profiling them with the NVIDIA Nsight Systems.
The interval of the CUDA stream in the profiled timeline is treated as a bubble when it is not occupied by any GPU kernel.
The bubble ratio is then defined as the proportion of all such bubbles in the compute stream.
Since \sysname{} has two active concurrent streams, we compute the bubble ratio for each stream and report their average as the final result.
Notably, the bubble ratio is a temporal metric and does not reflect how GPU kernels utilize the parallel GPU resources they occupy.

\sysname{} has a slightly higher bubble ratio ($7.7\%$ vs. $4.5\%$) due to its fine-grained kernel scheduling.
These extra bubbles occurs when the system is purely processing decode iterations and all prefill layers are completed.
Fortunately, these bubble do not degrade goodput, as there are no pending prefill launches and the decode iteration SLO is not violated.
The reported GPU utilization in \S\ref{eval:slo} also prove this.

\subsubsection{Preemptive Scheduling for Long Request}
\label{eval:preempt}
We evaluate the benefit of the bubble-less multiplex engine for preemptive scheduling by mixing requests from ShareGPT and LooGLE (50\% each).
Requests are generated at a rate of 0.5 per second following a Poisson process.
\autoref{fig:ttft-per-token-cdf} shows the CDF of TTFT per token with and without preemptive scheduling.
As shown, \sysname{} achieves a $1.96\times$ speedup on the $99\%${\it -ile} TTFT per token, demonstrating that it can also be configured to support more advanced SLO-aware scheduling policies.

\subsection{Overhead for Realizing PD-Multiplexing}\label{eval:overhead}
\mypar{Memory}
\sysname{} introduces some memory overhead by integrating GreenContext into existing serving systems.
Creating a group of green contexts requires only $4$MB, which is negligible compared to the total memory of modern GPUs.
However, integrating it with CUDA Graph incurs a $6.2\%$ overhead for both Llama-8B and Llama-70B on servers with 8 A100 or 8 H100 GPUs.
This arises because the serving system records kernel launches for each decode-phase batch size into a CUDA Graph, consuming extra GPU memory.
In \sysname{}, there are six partition configurations in total, and each decode-phase compute partition created by GreenContext adds memory usage for all recorded batch sizes.
Given the impressive performance gains of \sysname{}, this overhead is acceptable.

\mypar{Runtime}
\sysname{} splits the prefill phase into multiple prefill layers to enable bubble-less scheduling. This may introduce extra overhead due to fine-grained kernel launches.
We conduct an experiment to compare full prefill launching with layer-wise launching, where the prefill phase is split into the finest granularity.
Across various configurations with different batch sizes and context lengths, the total overhead remains within $1.5\%$.

%% file: section/discussion.tex
\section{Discussion}\label{sec:discuss}

\mypar{Generality of \sysname{}}
\sysname{} generalizes to accelerators that support intra-process spatial sharing with lightweight dynamic adjustment, such as GreenContext~\cite{nvidia2025grecontexts} on NVIDIA GPUs (supported since the Pascal architecture) and \texttt{hipExtStreamCreateWithCUMask()} on AMD GPUs.

\mypar{Contexts where \sysname{} excels}
\sysname{} targets scenarios with strict SLO guarantees (e.g., a decode-phase SLO below $100 ms$).
This is also the prevailing trend for achieving Model-as-a-Service in LLM serving.
In this setting, \sysname{} excels over existing works due to its efficient, fine-grained, and dynamic resource management between the prefill and decode phases.
When the SLO target is loose or absent, such as in offline serving, \sysname{} has no opportunity to outperform baselines such as chunked-prefill~\cite{zuoTamingSparsely} or NanoFlow~\cite{zhu_nanoflow_2025}.

\mypar{Large-scale deployment}
While \sysname{} is a single-instance optimization for high-goodput LLM serving, it can still benefit large-scale distributed deployments.
In such deployments, \sysname{} is complementary to disaggregated serving, as it optimizes each individual instance.
Specifically, low-utilization decode instances could be replaced with \sysname{} instances to exploit idle resources via spatially multiplexing prefill.
If prefill instance serves short requests—resulting in low utilization--or multiplexing decode on it does not violate TTFT SLO, it can be also utilized for higher efficiency.
However, when prefill instances consistently handle long requests (e.g., using chunked pipeline parallelism~\cite{qinMooncakeTrading}) or decode multiplexing violates TTFT SLO, \sysname{} offers limited benefit.

%% file: section/related_work.tex
\section{Related Work}
\mypar{Multiplexing in LLM Serving}
There are also prior works~\cite{WindServe,Tropical} that multiplex the prefill and decode phases in LLM serving.
WindServe~\cite{WindServe} multiplexes prefill and decode using a normal CUDA stream, which leads to uncontrollable contention.
It also does not address bubbles during scheduling.
Our prototype implementation of WindServe shows that, on ShareGPT, \sysname{} achieves a $1.61\times$ goodput improvement under a $50ms$ TBT SLO on an A100 with Llama-8B.
Tropical~\cite{Tropical} replaces the decode instance in disaggregated serving with temporally multiplexed prefill and decode.
It launches a full prefill only when sufficient slack exists.
When developing \sysname{}, we implemented an enhanced temporal-only variant that splits prefill into layers to fit small slacks.
It performs at least $20\%$ worse than \sysname{} because it cannot spatially leverage wasted resources.
There are also two similar community works~\cite{semi-pd}.
Semi-PD~\cite{semi-pd} utilizes MPS for multiplexing.
MPS enables \emph{inter-process} spatial sharing but requires process restarts to adjust SM allocations.
Semi-PD mitigates this by introducing a resident process and two additional inference engines, which adds significant complexity to existing frameworks.
Bullet~\cite{bulletserve} relies on \texttt{libsmctrl}~\cite{libsmctrl} to control SM allocation.
While Bullet claims that it can dynamically change the SMs allocated to each CUDA graph, our trials with Bullet’s open-sourced implementation show that the SM allocation for each CUDA graph does not change.
This also aligns with the claim in \texttt{libsmctrl}~\cite{libsmctrl} that it does not work with CUDA graph.

\mypar{Compute management in LLM serving}
There are two main approaches to compute management for improving system throughput under SLO constraints: disaggregation-based and fusion-based methods. On the one hand, DistServe~\cite{zhongDistServeDisaggregating} and Splitwise~\cite{patelSplitwiseEfficient} disaggregate LLM serving into separate prefill and decode instances, while LoongServe~\cite{wuLoongServeEfficiently} improves adaptability by enabling dynamic switching between the two at runtime. On the other hand, chunk-prefill~\cite{agrawalSARATHIEfficient} splits the prefill phase into chunks and fuses each chunk with a decode iteration for execution. However, disaggregation-based methods incur significant resource waste due to the coupled management of compute and memory, whereas fusion-based methods fail to fully maximize system throughput under SLO constraints. In contrast, our work decouples compute and memory management and maximizes goodput through spatial multiplexing.

\mypar{Memory management in LLM serving}
To improve system throughput in multi-turn or context-heavy LLM workloads, several systems propose memory management techniques. PagedAttention~\cite{kwonEfficientMemory} introduces a paged memory pool to enable KV cache reuse between prefill and decode phases. Parrot~\cite{linParrotEfficient} and SGLang~\cite{zhengEfficientlyProgramming} leverage context-aware caching to maximize reuse of KV segments across requests.  \sysname{} enhances these approaches by preserving memory sharing across phases and requests.

\mypar{Execution time modeling}
Performance modeling under spatial sharing is highly challenging, and prior efforts\cite{stratiOrionInterferenceaware,kimKschedulerDynamic,zhangLaiusLatency,zhangImprovingGPU} mainly focus on predicting interference for specific operators. GPUlet\cite{choi2022serving} uses linear regression with L1 cache utilization and DRAM bandwidth as input features to estimate performance interference among colocated operators. HSM\cite{hsm} and GDP\cite{gdp} also adopt linear regression based on low-level metrics in the simulator for operator slowdown prediction. 

\mypar{Compute partition techniques}
Existing GPU partitioning techniques can be broadly categorized into time-sharing and space-sharing approaches. Time-sharing is typically implemented via API remoting~\cite{chenBaymaxQoS,qcuda}. 
However, time-sharing alone is insufficient to meet \sysname{}’s requirements, as the prefill and decode phases already interleave in a time-sharing manner.
In contrast, NVIDIA provides MPS~\cite{nvidia_mps}, MIG~\cite{nvidia2025mig}, and GreenContext~\cite{nvidia2025grecontexts} for spatial sharing. MPS and MIG support inter-process spatial multiplexing, while GreenContext~\cite{nvidia2025grecontexts} enables intra-process spatial multiplexing with precise SM partition~\cite{nvidia2025stream}. \sysname{} builds on GreenContext to implement its PD multiplexing approach.

%% file: section/conclusion.tex

\section{Conclusion}

LLM services requires high goodput, yet existing serving systems struggle due to various deficiencies.
To address these issues, we present \sysname{}, an LLM serving framework with high goodput.
\sysname{} leverages a promising new serving paradigm, intra-GPU PD multiplexing, to achieve more flexible compute management for prefill and decode phases in LLM serving.
Experiments show that \sysname{} improves goodput by $2.2\times$ on average over state-of-the-art baselines.
Despite the notable performance improvement, \sysname{} also introduces a simple yet effective design for current LLM serving systems.
We plan to open-source \sysname{} after publication.

%

%% file: section/ae.tex
\appendix
\section{Artifact Appendix}

\subsection{Abstract}

\sysname{} is an LLM serving framework adopting intra-GPU prefill-decode multiplexing, which is built on the top of SGLang\cite{zhengEfficientlyProgramming}. We provide the source code of \sysname{} and scripts to reproduce comparison of chunked-prefill. This appendix includes instructions for reproducing similar data in \autoref{fig:other-gpus-llms} and \autoref{fig:e2e-synthetic}.

\subsection{Artifact check-list (meta-information)}

{\small
\begin{itemize}
  \item {\bf Model: } CodeLlama-34b-Instruct-hf.
  \item {\bf Data set: } ShareGPT \cite{sharegpt_2023} and LooGLE \cite{li2024looglelongcontextlanguagemodels}.
  \item {\bf Hardware: } 
  
  NVIDIA H200 NVL (140 GB, 132 SMs) 

  NVIDIA driver: 580.65.06 (must be greater than 570)
  \item {\bf Experiments: } This appendix provides instrucitons for comparing $99\%${\it -ile} TTFT and $99\%${\it -ile} TBT \sysname{} between \sysname{} and chunked-prefill under various workload.
  \item {\bf Metrics: }
  $99\%${\it -ile} TTFT,
  $99\%${\it -ile} TBT
  \item {\bf Output: } Jsonl files containing metrics from \sysname{} and chunked-prefill with different chunk size.
  \item {\bf How much disk space required (approximately)?: } Approximately 200GB
  \item {\bf How much time is needed to prepare workflow (approximately)?: } About 10 minutes to build from source code.
  \item {\bf How much time is needed to complete experiments (approximately)?: } About 2 hours for ShareGPT workload and 4 hours for LooGLE workload.
  \item {\bf Publicly available?: } Yes.
\end{itemize}
}

\subsection{Description}

\subsubsection{How to access}
The source code of \sysname{} is available for download on Zenodo: \url{https://zenodo.org/records/18062118}.
The pre-built Docker image can be found in: \url{https://hub.docker.com/layers/combathhhhhh/pdmux/sglpr_torch2.6_bench}

\subsubsection{Hardware dependencies}
Requires an x86-64 Linux host with at least 200 GB of free disk space, and an NVIDIA H200 NVL GPU (140 GB, 132 SMs).
\subsubsection{Software dependencies}
NVIDIA driver: 580.65.06 (must be greater than 570).
\subsubsection{Data sets}
ShareGPT: chatbot tasks, with an average input length of 226 and average output length of 195.

LooGLE: long-context understanding tasks, with an average input length of 30k and average output length of 15.
\subsubsection{Models}
CodeLlama-34b-Instruct-hf.
\subsection{Installation}



  

Please follow the instructions below, which are adapted from our GitHub repository (\url{https://github.com/ykcombat/sglang/tree/slo_config}):

{\small
\begin{verbatim}
# 1. Clone the repository and switch to the slo_config branch
git clone https://github.com/ykcombat/sglang.git
cd sglang
git checkout slo_config

# 2. Build SGLang
pip install --upgrade pip
pip install -e "python"

\end{verbatim}
}
\subsection{Experiment workflow}
Our experiments focus on comparison between \sysname{} and chunked-prefill.
\begin{enumerate}
    \item Download the required LLM model(CodeLlama-34b-Instruct-hf) to /workspace/data.
    \item Start \sysname{} or chunked-prefill server. You can change the environment virable \$CHUNK\_SIZE to start chunked-prefill server with different token budgets.
    {\small
    \begin{verbatim}
    # 1. Start MuxWise Server
    ./start_pdmux.sh
    
    # 2. Start Chunked-prefill Server
    ./start_chunk.sh
    \end{verbatim}
    }
    \item Start evaluating in another terminal.
    {\small
    \begin{verbatim}
    # 1. Evaluate MuxWise on ShareGPT and LooGLE
    ./bench_pdmux.sh
    
    # 2. Evaluate Chunked-prefill on ShareGPT and LooGLE
    ./bench_chunk.sh
    \end{verbatim}
    }
\end{enumerate}
\subsection{Evaluation and expected results}
When all experiments done, you will obtain jsonl files containing detailed metrics under /workspace/sglang. To visualize the results, run plot.ipynb; this will generate figures similar to the reference plot provided at /workspace/ sglang/H200\_result.png. Please note that results may vary depending on the specific hardware used. You can refer to \url{https://github.com/ykcombat/sglang/blob/slo_config/README.md} for more information.
\subsection{Notes}
When serving different workloads, different configurations are used, which can be found in our repository.
